\documentclass{aa}  

\usepackage{graphicx}
\usepackage{xcolor}
\usepackage{multicol}
\usepackage{txfonts}
\def\PLUTO{{\sc pluto}}

\newcommand{\casa}{Cas~A}
\newcommand{\sal}{}

\begin{document} 


    \title{Deciphering the "Green Monster" in Cassiopeia A: Puncturing and sculpting a heterogeneous circumstellar shell}




\author{S.\,Orlando\inst{1}
      \and H.-T.\,Janka\inst{2}
      \and D.\,Milisavljevic\inst{3}
      \and I.\,De Looze\inst{4}
      \and T.\,Temim\inst{5}
      \and R.\,Fesen\inst{6}
      \and B.-C.\,Koo\inst{7}
      \and \\M.\,Miceli\inst{8,1}
      \and F.\,Bocchino\inst{1}
  }

\institute{INAF -- Osservatorio Astronomico di Palermo, Piazza del Parlamento 1, I-90134 Palermo, Italy\\
\email{salvatore.orlando@inaf.it}
\and Max-Planck-Institut f\"ur Astrophysik, Karl-Schwarzschild-Str. 1, D-85748 Garching, Germany
\and Department of Physics and Astronomy, Purdue University, 525 Northwestern Avenue, West Lafayette, IN 47907, USA
\and Sterrenkundig Observatorium, Ghent University, Krijgslaan 281 - S9, B-9000 Gent, Belgium
\and Princeton University, 4 Ivy Ln, Princeton, NJ 08544, USA
\and Department of Physics and Astronomy, 6127 Wilder Laboratory, Dartmouth College, Hanover, NH 03755, USA
\and Department of Physics and Astronomy, Seoul National University, Gwanak-ro 1, Gwanak-gu, Seoul, 08826, Republic of Korea
\and Dip. di Fisica e Chimica, Universit\`a degli Studi di Palermo, Piazza del Parlamento 1, 90134 Palermo, Italy
             }

   \date{Received xxx, xxx; accepted xxx, xxx}

 
  \abstract
   {JWST observations of Cassiopeia A have revealed the "Green Monster" (GM), a pockmarked region of shocked circumstellar medium (CSM) featuring circular holes surrounded by bright rings. The physical origin of these structures is actively debated between post-shock sculpting by ejecta fingers or pre-shock puncturing by fast-moving knots (FMKs).}
   {We quantitatively test the physical viability of the FMK-driven scenario against observed GM morphologies and kinematics, and establish observational diagnostics to distinguish among the competing formation mechanisms.}  
   {We perform three-dimensional hydrodynamic simulations of isolated FMKs interacting with a dense circumstellar shell, followed by the passage of the supernova remnant's forward shock. A range of FMK properties and shell densities is explored, comparing the resulting hole-ring systems directly with JWST observational constraints.}
   {In the FMK-driven scenario, holes and rings are formed by bow shocks that precede the knots, in contrast to the post-shock scenario, where cavities result from ejecta displacing material that has already been processed by the forward shock. While primary FMKs can reproduce qualitative GM-like features, they generally produce hole-ring systems that are significantly larger than the $1\arcsec-3\arcsec$ (namely $\sim 0.016-0.048$~pc at the distance of $\approx 3.4$~kpc of \casa) features observed by JWST unless the knots are slow ($\lesssim 8000$~km~s$^{-1}$) and the shell is dense ($n_{\rm sh} \gtrsim 200$~cm$^{-3}$). Furthermore, these structures are  transient, typically distorting within $30-60$ years after forward-shock passage.} 
   {The GM's diverse hole-ring systems and complex kinematics cannot be attributed to a single idealized scenario, but can be instead explained by the simultaneous action of three mechanisms operating within a structured, heterogeneous CSM. While large cavity relics are carved by early-stage primary FMKs, compact pristine rings are produced by recent "first contact" punctures of secondary knots from fragmented ejecta fingers, and older structures are continuously sculpted by long-term post-shock interactions with large-scale ejecta fingers.}


\titlerunning{Deciphering the "Green Monster" in Cassiopeia A}

\authorrunning{S. Orlando et~al.}

   \maketitle
%

\section{Introduction}

Cassiopeia A (\casa) is a young ($\sim 350$~yr; \citealt{2001AJ....122..297T, 2006ApJ...645..283F}), nearby (3.4~kpc; \citealt{1995ApJ...440..706R}) supernova remnant (SNR) providing a crucial window into the physics of core-collapse supernovae (SNe) and their interaction with the surrounding medium. Recent James Webb Space Telescope (JWST) observations (\citealt{2024ApJ...965L..27M}) have revealed unprecedented ejecta filaments and previously unseen circumstellar medium (CSM) features. These data offer a unique opportunity to link the remnant's morphology to the physical processes of the explosion and the final evolutionary stages of its progenitor.

Newly revealed structures include a web-like network of filaments interior to the reverse shock, enriched in intermediate-mass elements (O, Ne, Mg) and resolved to $\sim 0.01$~pc (\citealt{2024ApJ...965L..27M}; Dickinson et al., in prep.). Simulations of neutrino-driven explosions (\citealt{2017ApJ...842...13W, 2021A&A...645A..66O}) indicate these filaments provide a "fossil record" of early explosion dynamics. They encode imprints of neutrino-heated bubbles, hydrodynamic (HD) instabilities during shock propagation in the stellar interior, and the post-breakout Ni-bubble effect (\citealt{2025A&A...696A.108O}). Collectively, these mechanisms shaped the current ejecta morphology, preserving signatures of the earliest SN stages.

Another remarkable JWST discovery is the so-called "Green Monster" (GM), a region of shocked circumstellar material with a striking pockmarked morphology, characterized by almost regular circular holes surrounded by bright rings (\citealt{2024ApJ...976L...4D}). The GM may represent the relic of an asymmetric dense circumstellar shell that has interacted with \casa\ within the last $100-150$~years, potentially explaining the unusual properties of the reverse shock observed in X-ray and optical bands (\citealt{2022ApJ...929...57V, 2022A&A...666A...2O, 2025ApJS..278...17F}). 

Two main scenarios have been proposed to explain the origin of the GM’s rings and holes (\citealt{2024ApJ...976L...4D}). In the first scenario (hereafter referred to as the post-shock scenario), Rayleigh–Taylor (RT) fingers and dense ejecta clumps developing at the contact discontinuity in the remnant penetrate the circumstellar shell after it has been swept by the forward shock. As these structures advance, they carve holes in the shell and pile up material along their boundaries, naturally giving rise to dense, ring-like features (\citealt{2022A&A...666A...2O, 2025A&A...696A.188O}). 

While this mechanism successfully reproduces the GM's morphological traits \citep{2025A&A...696A.188O}, it predicts shell velocities of approximately $-2000$~km~s$^{-1}$ (i.e., toward the observer). These high velocities align with the blueshifted values inferred from Chandra X-ray observations \citep[$\approx -2300$~km~s$^{-1}$;][]{2024ApJ...964L..11V}, yet they stand in stark contrast to the near-rest-frame radial velocities of $-50$ to $0$~km~s$^{-1}$ measured in the bright infrared rings by JWST \citep{2024ApJ...976L...4D}. This discrepancy suggests that the infrared-emitting dust and gas trace a significantly slower, denser component of the CSM than the hot, X-ray-emitting plasma.

The second scenario (hereafter the FMK-driven scenario) involves fast-moving knots (FMKs), dense, metal-rich clumps of ejecta, largely devoid of H and He emission, traveling at several thousand kilometers per second (\citealt{2001AJ....122.2644F, 2006ApJ...636..859F, 2006ApJ...645..283F, 2013ApJ...772..134M, 2017ApJ...837..118L, 2018ApJ...866..139K, 2026arXiv260209175J}). FMKs trace the most energetic and asymmetric components of the explosion and are often observed extending beyond the main remnant outline. FMKs have been observed in other core-collapse SNRs (e.g. \citealt{2003A&A...400..203B, 2012A&A...541A.152B}) and even in Type Ia SNRs (e.g., \citealt{2024A&A...684A..68G}).

\begin{figure}
   \centering
   \includegraphics[width=0.47\textwidth]{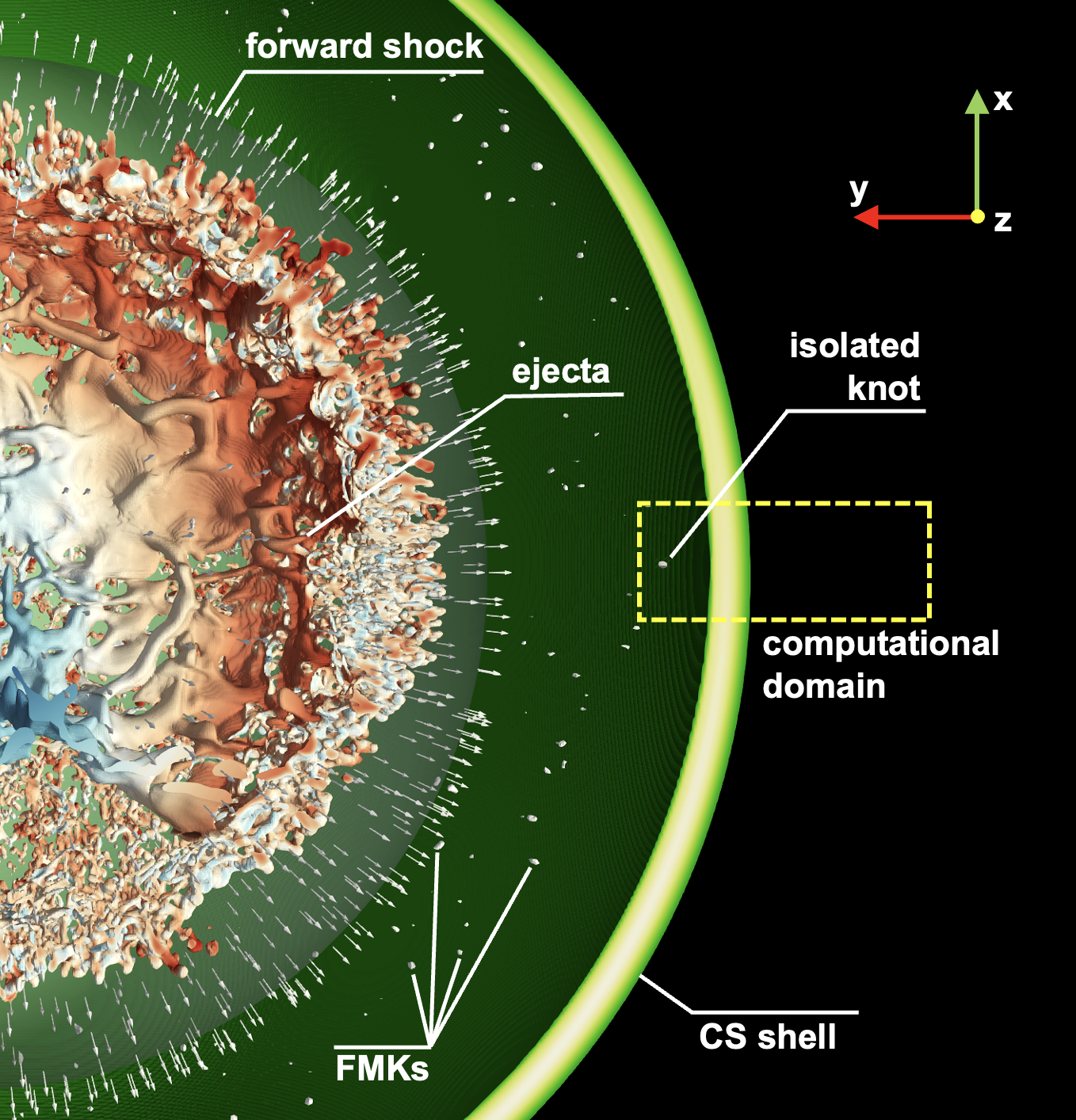}
    \caption{Schematic illustration of the FMK-driven scenario. The SNR expands into a CSM containing an asymmetric, dense shell (green). White arrows indicate the outward propagation of the forward shock, while fast-moving knots (FMKs) of ejecta (white bullets), positioned ahead of the shock front, approach the shell. The dashed box delineates the cross-section of the 3D HD computational domain, which focuses on a localized portion of the shell and an isolated FMK. This setup is designed to investigate the knot-shell interaction and the subsequent morphological evolution of the resulting structures following shock passage.}
   \label{fig1}%
   \end{figure}

\cite{2024ApJ...976L...4D} proposed that, if FMKs puncture the dense circumstellar shell before the arrival of the forward shock (see Fig.~\ref{fig1}), they can pre-shape a pockmarked structure. The subsequent impact of the primary SN blast wave would then shock the material, potentially explaining the observed warm dust temperatures (130 to 300 K; \citealt{2024ApJ...976L...4D}) and high gas temperatures ($5\times 10^{6}$ to $10^{7}$~K; see also \citealt{2024ApJ...964L..11V}). However, this model faces a significant dynamical challenge: the interaction with the forward shock would likely distort or destroy the delicate circular morphology of the rings unless they possess incredibly high densities. While the measured line broadening of $100-200$~km~s$^{-1}$ (\citealt{2024ApJ...976L...4D}) confirms the GM is currently in a shocked state, it remains unclear whether pre-formed rings could survive such a major shock interaction without devolving into the irregular structures typical of the rest of the remnant.

Establishing which scenario dominates the formation of the GM is important for understanding both the explosion dynamics and the circumstellar environment of \casa. If RT fingers and ejecta clumps sculpt the shell only after the passage of the forward shock, then the observed GM morphology primarily reflects post-shock instabilities developing at the contact discontinuity. Conversely, if FMKs puncture the shell prior to the arrival of the shock, the resulting structures retain the imprint of the most energetic and asymmetric components of the explosion, thereby providing a more direct probe of the early SN anisotropies.

Identifying the dominant mechanism constrains CSM properties, informs multi-wavelength interpretation, and tests 3D core-collapse SN models (e.g., \citealt{2015A&A...577A..48W, 2017ApJ...842...13W, 2017hsn..book.1095J, 2021Natur.589...29B, 2025ApJ...982....9V}). Ultimately, the GM serves as a high-resolution tracer of the large-scale CSM architecture. Its quasi-stationary nature points to a massive, potentially equatorial mass-loss episode consistent with a stripped-envelope binary progenitor. On a broader scale, the spatially resolved kinematics of this multi-density structure provide a vital local laboratory for the wider astrophysical community. For instance, by mapping these real-time shock-CSM interactions, these observations may offer physical insights needed to decode the complex, unresolved line profiles obtained via flash spectroscopy of strongly interacting extragalactic transients, such as Type IIn supernovae (e.g., \citealt{2023ApJ...954L..42J, 2023ApJ...956...46S}).

In this work, we investigate the FMK-driven scenario using three-dimensional (3D) HD simulations that follow the interaction of individual ejecta knots with the circumstellar shell, as well as the subsequent impact of the forward shock on the punctured shell. We explore a range of knot properties and shell densities to capture the diversity observed in \casa. By comparing the resulting morphologies with JWST observations and with the predictions of the post-shock scenario (\citealt{2025A&A...696A.188O}), we aim not only to assess the relative plausibility of the two mechanisms, but also to identify observational diagnostics capable of distinguishing the dominant formation channel of the observed hole–ring systems in the GM.

The paper is structured as follows. In Sect.~\ref{sec:model}, we describe the numerical setup adopted for our 3D HD simulations. Sect.~\ref{sec:results} presents the HD outcomes, while Sect.~\ref{sec:holes_rings} compares these simulated features directly with the observed properties of the GM in \casa. In Sect.~\ref{sec:discussion_multiphase}, we examine the potential evidence pointing to the GM as a multiphase, heterogeneous architecture. Finally, Sect.~\ref{sec:summary} summarizes our primary findings and outlines their broader astrophysical implications. Appendix~\ref{app:code} details the model's implementation within the \textsc{pluto} code, and Appendix~\ref{app:holes_evol} provides supplementary analysis on the long-term temporal evolution of the hole--ring systems.

\section{Three-dimensional hydrodynamic modeling}
\label{sec:model}

We model the interaction of an isolated FMK with the dense circumstellar shell surrounding \casa\ \citep{2022A&A...666A...2O}. The scenario follows a two-stage progression: (i) the FMK, traveling ahead of the remnant's outer boundary, punctures the unperturbed shell, and (ii) the remnant's forward shock subsequently collides with this already-punctured structure, shaping the resulting hole and bright ring. To resolve the relevant dynamics, we simulate a localized region encompassing the FMK and a portion of the shell (see the dashed box in Fig.~\ref{fig1}). For a more direct comparison with the results of \cite{2025A&A...696A.188O} and with the structures observed in the GM by JWST (\citealt{2024ApJ...976L...4D}), the model is oriented such that the FMK propagates toward the observer along the $y$-axis, coincident with the line of sight (LoS), with Earth located in the negative $y$-direction.

The FMK is initialized in pressure equilibrium with the surrounding medium and located between the forward shock and the shell, allowing it to reach and puncture the shell before the shock arrives. It is modeled as an initial spherical clump, fully characterized by its density ($n_{\rm fmk}$), radius ($r_{\rm fmk}$), and velocity ($u_{\rm fmk}$). The forward shock is initialized to match the properties of the \casa\ forward shock at an age of $100-150$~years, as inferred from model W15-IIb-sh-MHD+dec-rl-hr of \citet{2025A&A...696A.188O}. In this model, the forward shock encounters the circumstellar shell at $t \approx 200$~years after the SN event. 

Following \citet{2022A&A...666A...2O}, both the knot and the remnant expand through the spherically symmetric wind of the progenitor star, with a density profile scaling as $r^{-2}$, where $r$ is the radial distance from the explosion center. The wind density is set to $n_{\rm w} = 0.8~\mathrm{cm^{-3}}$ at $r_{\rm fs} = 2.5$~pc, corresponding to the nominal current outer radius of the remnant at a distance of $\sim 3.4$~kpc, consistent with post-shock wind densities inferred from observations of \casa\ \citep{2014ApJ...789....7L}. The CSM also includes a dense shell embedded within the wind, located at a radial distance of $\sim 1.5$~pc from the explosion center \sal{and with a thickness of $\sigma \approx 0.05$~pc}; for a detailed description of the shell implementation, we refer to \citet{2022A&A...666A...2O}. Both the wind and the shell are expected to have characteristic expansion velocities of order tens of km s$^{-1}$, which are negligible compared to the expansion velocity of the remnant. We therefore assume, for simplicity, that the wind and shell are at rest. The CSM, comprising both the wind and the shell, is assumed to have standard cosmic abundances.

\begin{table*}
\caption{Explored parameters and initial conditions for isolated FMK.}
\label{tab1}
\begin{center}
\begin{tabular}{lllllll}
\hline
\hline
Run & Model name              & $r_{\rm fmk}$  & $u_{\rm fmk}$      & $t_0$   & $\Delta t$   & $n_{\rm sh}$\\ 
    &                         & [$10^{16}$~cm] & [$10^3$~km~s$^{-1}$] & years & years        & [cm$^{-3}$] \\ \hline
1   & FMK–R1.5–U8–T100-SH40   & 1.5            & 8                  & 100     &  100         & 40 \\
2   & FMK–R1.5–U8–T150-SH40   & 1.5            & 8                  & 150     &  50          & 40 \\
3   & FMK–R5–U8–T150-SH40     & 5              & 8                  & 150     &  50          & 40 \\
4   & FMK–R1.5–U6–T100-SH40   & 1.5            & 6                  & 100     &  100         & 40 \\
5   & FMK–R1.5–U10–T130-SH40  & 1.5            & 10                 & 130     &  70          & 40 \\
6   & FMK–R1.5–U15–T150-SH40  & 1.5            & 15                 & 150     &  50          & 40 \\
7   & FMK–R1.5–U8–T130-SH200  & 1.5            & 8                  & 130     &  70          & 200 \\
8   & FMK–R1.5–U8–T150-SH200  & 1.5            & 8                  & 150     &  50          & 200 \\
9   & FMK–R1.5–U15–T150-SH200 & 1.5            & 15                 & 150     &  50          & 200 \\
\hline
\end{tabular}
\end{center}
\end{table*}

The dynamics of the system is described by solving the time-dependent HD equations of mass, momentum, and energy conservation. \sal{The plasma is described by an ideal-gas equation-of-state with an adiabatic index $\gamma =5/3$.} The simulations include radiative cooling from an optically thin plasma, with losses $\Lambda(T_{\mathrm e},\tau,Z)$ computed locally in each cell as a function of the electron temperature $T_{\mathrm e}$, the ionization age $\tau$, and the chemical abundances $Z$. This treatment ensures that cooling reflects the actual physical and chemical state of the plasma. The numerical implementation of the radiative losses is described in detail in \citet{2025A&A...696A.188O}. All simulations were performed with the \textsc{pluto} code, a modular Godunov-type framework for astrophysical fluid dynamics \citep{2012ApJS..198....7M}. In Appendix \ref{app:code}, we describe the details of the implementation of the model and the numerical grid adopted.

We explored a broad region of parameter space by varying the radius and velocity of the knot. The adopted ranges for $r_{\rm fmk}$ and $u_{\rm fmk}$ are consistent with those inferred for the FMKs in \casa. Specifically, we considered models with $r_{\rm fmk}$ in the range $(1.5-5)\times10^{16}$~cm and $u_{\rm fmk}$ between 6000 and 15000~km~s$^{-1}$ (e.g., \citealt{2001AJ....122.2644F, 2006ApJ...645..283F, 2008ApJS..179..195H, 2023ApJ...953..131K, 2026arXiv260209175J}). The knot density was fixed at $n_{\rm fmk}=10^{4}$~cm$^{-3}$ (according to the masses estimated by \citealt{2002A&A...390..327B}) to ensure that the knot survives the interaction with the shell and effectively pierces it, producing a localized opening. We also explored different epochs for the knot–shell interaction, corresponding to different times at which the FMK impacts the shell prior to the arrival of the forward shock. 

For most models, we adopted the shell configuration constrained by \cite{2022A&A...666A...2O}, which yields a maximum shell density of $n_{\rm sh}=40$~cm$^{-3}$ in the interaction region considered here. However, since this density constraint was primarily derived to reproduce the properties of the reverse shock in the western region of the remnant (see \citealt{2022A&A...666A...2O}), we cannot exclude that the local shell density associated with the GM may differ significantly. To assess how the dynamical and morphological evolution of the hole–ring system depends on the shell density, we therefore also explored models in which the FMK interacts with a denser shell, with a maximum density of $n_{\rm sh}=200$~cm$^{-3}$.

A summary of all simulated models is reported in Table~\ref{tab1}. The naming convention adopted throughout this work is FMK-R$r_{\rm fmk}$-U$u_{\rm fmk}$-T$t_0$-SH$n_{\rm sh}$, where $r_{\rm fmk}$ is the knot radius in units of $10^{16}$~cm, $u_{\rm fmk}$ is the knot velocity in units of $10^{3}$~km~s$^{-1}$, $t_0$ is the epoch of knot impact on the shell expressed in years since the explosion, and $n_{\rm sh}$ is the maximum shell density in $\mathrm{cm^{-3}}$. It is worth noting that while our underlying global model assumes a specific radial distance for the shell and a nominal forward-shock arrival time of $t \approx 200$~years post-SN, our simulations focus on a highly localized region where the FMK penetrates the shell. Since the pre-shock structure and exact spatial position of the shell remain unconstrained, the actual arrival time of the forward shock can vary significantly across different regions of the remnant, rendering the absolute arrival epoch secondary. The physically critical parameter governing the morphological and kinematic evolution of the hole-ring networks is instead the precise local time interval, $\Delta t$, elapsed between the initial impact of the FMK and the subsequent arrival of the forward shock at the shell boundary. Consequently, this local delay time, $\Delta t$ (expressed in years), is explicitly reported in Table~\ref{tab1}.

\section{Hydrodynamic evolution}
\label{sec:results}

We first analyzed the interaction of the FMK with the unperturbed circumstellar shell surrounding \casa\ (Sect.~\ref{sec:shell_evol}). Figure~\ref{fmk_strcuct} shows the evolution of our reference model FMK-R1.5-U8-T100-SH40 (Table~\ref{tab1}), illustrating the formation of shell structures induced by the FMK. The subsequent evolution of these structures under the remnant’s forward shock is discussed in Sect.~\ref{sec:shock-shell}. Other models show qualitatively similar behavior, indicating that such features naturally arise from FMK–shell interactions over the explored parameter range.


\subsection{Formation of holes and rings in the shell}
\label{sec:shell_evol}

   \begin{figure*}
   \centering
   \includegraphics[width=0.75\textwidth]{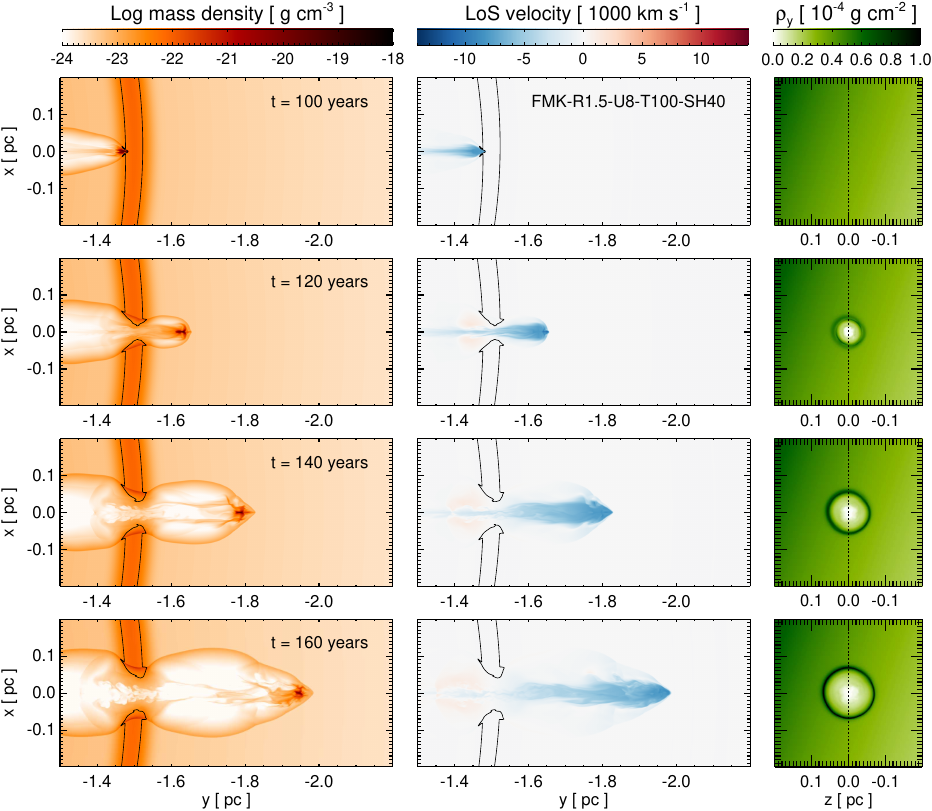}
   \caption{Interaction between a FMK and the unperturbed circumstellar shell around \casa\ in model FMK-R1.5-U8-T100-SH40, shown at the epochs indicated in the upper-right corner of the left panels. The model is oriented with Earth along the negative $y$-axis, so that the FMK moves toward the observer along the LoS. Left: mass-density cross-sections (logarithmic scale) in a plane containing the LoS; the contour outlines the shell. Center: corresponding LoS velocity maps. Right: 3D volumetric renderings of the shell density integrated along the LoS within the selected region, as viewed from Earth. The vertical dashed line marks the position of the cross-sections shown in the left and center panels. The interaction excavates a hole in the shell surrounded by a dense, bright ring. \sal{In all models, the forward shock reaches the shell at $t=200$ yr; the time to the shock impact is therefore obtained by subtracting 200 yr from the labeled time.}
   }
   \label{fmk_strcuct}%
   \end{figure*}

In the scenario considered here, FMKs begin interacting with the circumstellar shell several decades ($50-100$~years; see $\Delta t$ in Table~\ref{tab1}) before the arrival of the forward shock. In the reference model, this interaction starts at approximately $t \approx 100$~yr after the SN explosion, that is, about a century before the forward shock reaches the shell. As the FMK penetrates the dense shell, it excavates a low-density cavity (the hole) that rapidly expands \sal{over a timescale of $\sim 20$~yr}, reaching a size of $\sim 0.08$~pc ($\sim 2.5\times10^{17}$~cm, corresponding to $\sim 5\arcsec$ at a distance of $3.4$~kpc), far exceeding the initial knot radius of $\sim 1.5\times10^{16}$~cm (see second-row panels in Fig.~\ref{fmk_strcuct}). This rapid enlargement is not driven by direct mechanical displacement of shell material by the knot itself, but instead results from the bow shock generated ahead of the FMK as it propagates supersonically through the CSM. The bow shock compresses and accelerates the shell gas laterally, producing an expanding cavity. A bright ring naturally forms at the cavity boundary, corresponding to the dense shell material accumulated and compressed at the bow shock front. The bow shock continues to expand even after the FMK has traversed the shell, further enlarging both the cavity and the surrounding high-density ring. Sixty years after the onset of the interaction, the cavity–ring system in the reference model reaches a diameter of $\sim 0.14$~pc ($\sim 4.3\times10^{17}$~cm, corresponding to $\sim 8.5\arcsec$ at the distance of \casa, see Fig.~\ref{fmk_strcuct}). 

The size and morphology of the hole–ring system are highly sensitive to both the properties of the FMK and the local shell density. Faster knots (e.g., FMK–R1.5–U10–T130–SH40 and FMK–R1.5–U15–T150–SH40 in Table~\ref{tab1}) drive stronger bow shocks, producing larger cavities more rapidly and generating denser, more pronounced rims, while larger knots (e.g., FMK–R5–U8–T150–SH40) inject more momentum into the shell, accelerating the lateral expansion of the cavity. At comparable evolutionary times, faster or larger knots consistently produce wider holes and more prominent rings. The shell density further regulates this evolution: in models with a shell five times denser than the reference case (e.g., FMK–R1.5–U8–T120–SH200), the transmitted shock propagates more slowly, limiting lateral expansion and delaying hole formation. Consequently, at similar times, cavities in these denser-shell models are smaller than those in the reference case with $n_{\rm sh}=40$~cm$^{-3}$. Overall, the final size and brightness of the cavity–ring system reflect the combined influence of FMK velocity and size and the local shell conditions. In this sense, the FMK effectively pre-sculpts the shell, establishing the structures that will later be overtaken and modified by the forward shock.

It is interesting to note that the physical mechanism shaping holes and rings in the FMK-driven scenario is fundamentally different from that in the post-shock scenario \citep{2025A&A...696A.188O}. In the post-shock framework, holes arise from the displacement of shell material that has already been processed by the forward shock. Dense ejecta fingers and clumps, formed via RT instabilities at the contact discontinuity, penetrate the shocked shell, carving out cavities. The surrounding rings consist of shell material compressed and accumulated at the edges of these ejecta structures. In this scenario, the hole size provides a direct measure of the characteristic width of the ejecta fingers. As shown by \citet{2025A&A...696A.188O}, reproducing the relatively small sizes of the observed holes requires efficient radiative cooling within the ejecta, which enhances condensation, increases their density, and reduces both the thickness and temperature of the fingers.

   \begin{figure*}
   \centering
   \includegraphics[width=0.75\textwidth]{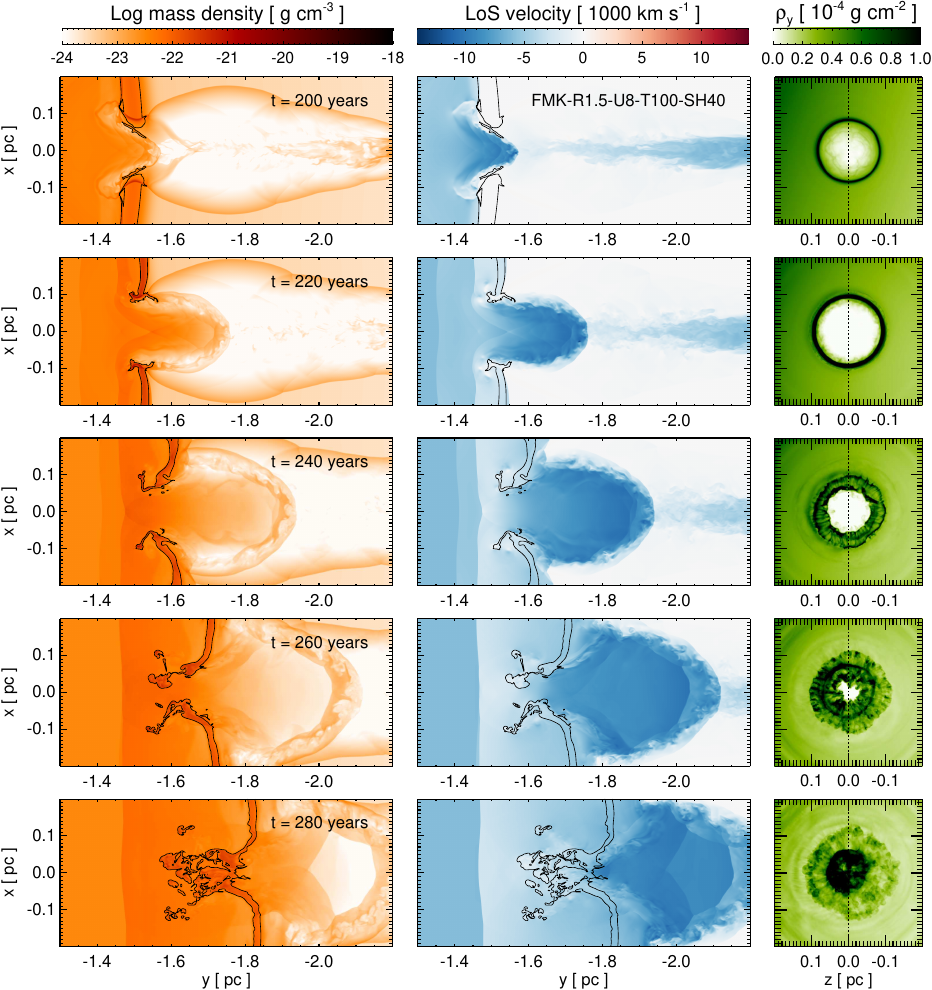}
   \caption{Same as in Fig.~\ref{fmk_strcuct}, but showing the subsequent interaction of the remnant’s forward shock, expanding from left to right, with the shell already pierced by the FMK in model FMK-R1.5-U8-T100-SH40. Times are indicated in the upper right corner of the left panels\sal{; the time since/to the shock impact is obtained by subtracting 200 yr from the labeled time.}}
   \label{fmk_evolution}%
   \end{figure*}

In contrast, the FMK-driven scenario explored here is governed by bow-shock dynamics: supersonic FMKs carve cavities in the shell, while the surrounding rings trace the compressed and displaced material swept up by the bow shock. The size of cavities and rings depends on the bow-shock strength and shell density. For an FMK velocity $u_{\rm fmk}$, the bow-shock Mach number, $M_{\rm sh} \sim u_{\rm fmk}/c_s$ (where $c_s$ is the local sound speed in the shell) determines the shock compression and lateral expansion of the shocked shell. Thus, the cavity size directly reflects the momentum and kinetic energy transferred by the FMK and the efficiency of the bow-shock–driven displacement of shell material.



\subsection{Interaction of the forward shock with the pierced shell}
\label{sec:shock-shell}


   \begin{figure*}
   \centering
   \includegraphics[width=0.85\textwidth]{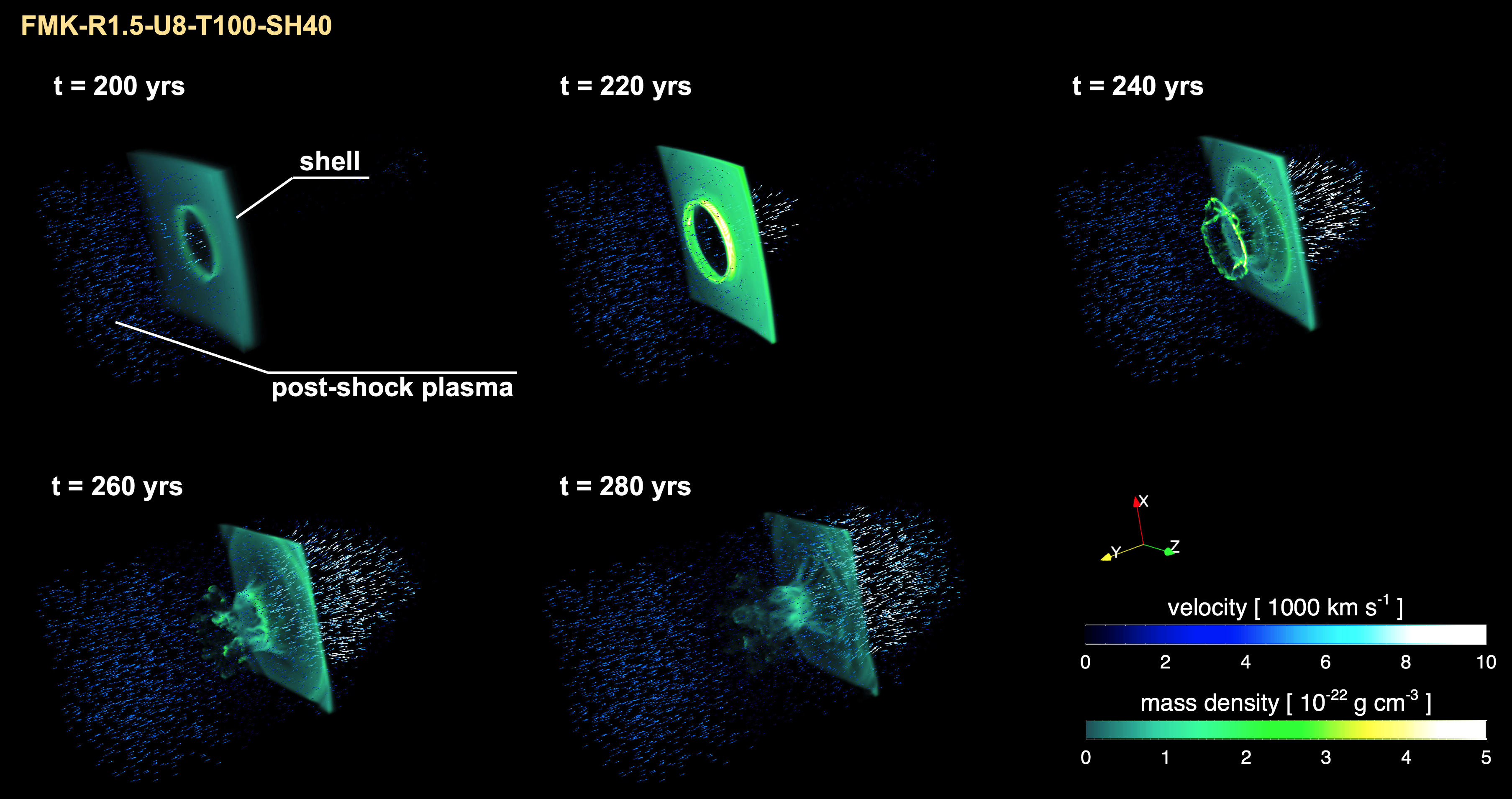}
   \caption{Evolution of the pierced circumstellar shell after the passage of the forward shock in model FMK-R1.5-U8-T100-SH40 at the same epochs shown in Fig.~\ref{fmk_evolution}\sal{; the time since shock impact is obtained by subtracting 200 yr from the labeled time.} The panels show 3D volume renderings of the shell (green; color scale at lower right), with arrows indicating the local velocity field. Arrow size and color (blue; lower-right scale) represent the velocity magnitude. The sequence illustrates the shell reshaping, the collapse of the dense ring, and the filling of the cavity left by the FMK.}
   \label{hole_disrupt}%
   \end{figure*}

Before reaching the shell, the forward shock is perturbed by the low-density wake left by the FMK (e.g., lower panels of Fig.~\ref{fmk_strcuct}), where its velocity locally increases and mildly distorts the remnant outline \sal{(see Appendix~\ref{app:holes_evol} and Fig.~\ref{fig:pre_impact})}. Upon encountering the FMK-excavated cavity, the shock accelerates further, producing a rarefied, fast flow through the hole (upper panels of Fig.~\ref{fmk_evolution}). Conversely, interaction with the unperturbed shell and the dense surrounding ring causes shock deceleration and compression, generating transmitted shocks into the shell and reflected shocks propagating back into the shocked wind. The resulting 3D flow (Fig.~\ref{hole_disrupt}) develops shear and vorticity along the ring boundaries due to velocity gradients and oblique shock interactions.


Over time, the forward shock reshapes the cavity initially carved by the FMK (see the time sequence in Figs.~\ref{fmk_evolution} and \ref{hole_disrupt}). The low-density cavity is gradually filled by shocked circumstellar material flowing inward along the sides of the ring. Meanwhile, the ring itself, composed of compressed shell material, propagates more slowly than the surrounding shell because its higher density increases its inertia and slows its response to the forward shock (Fig.~\ref{hole_disrupt}). As a result, the dense ring lags behind the ambient shell and is eventually overtaken by the post-shock flow driven by the remnant’s forward shock (see the upper-right panel in Fig.~\ref{hole_disrupt}).

When overtaken by the forward shock, the ring experiences strong compression and becomes highly susceptible to HD instabilities. The impulsive acceleration of the ring triggers Richtmyer–Meshkov instabilities, while velocity shear along its boundaries and the oblique incidence of the shock enhance vorticity and turbulence. These processes promote fragmentation of the ring and generate small-scale density inhomogeneities, as illustrated in the lower panels of Fig.~\ref{hole_disrupt}. In models with a high shell density (e.g., FMK–R1.5–U8–T120–SH200), the hole–ring system initially appears more stable; however, in these cases, efficient radiative cooling drives local thermal instabilities, that contribute in perturbing the hole-ring system. As this evolution proceeds, the FMK-carved low-density cavity is rapidly erased, and the fragmented ring collapses inward under the combined effects of shock compression and converging post-shock flows. The ultimate outcome is a compact, bright, high-density central structure (see the right panels in Fig.~\ref{fmk_evolution}), representing the final stage of the FMK-driven hole after it has been overtaken and processed by the remnant’s forward shock.

The timescale over which the hole–ring system evolves is set by the relative velocities and characteristic sizes of the FMK, the forward shock, and the local shell density. For the range of parameters explored in our models, FMK velocities $u_{\rm FMK} \sim 6000-15000~\mathrm{km~s^{-1}}$, FMK radii $r_{\rm FMK} \sim (1.5-5)\times10^{16}$~cm, and forward-shock speeds $u_{\rm FS} \sim 6000~\mathrm{km~s^{-1}}$ (e.g., \citealt{2022ApJ...929...57V}), the forward shock hits the FMK-carved cavity within a few decades of the knot’s passage through the shell. In models with higher shell density (e.g., $n_{\rm sh}=200~\mathrm{cm^{-3}}$), the transmitted shock propagates more slowly, and the disruption of the hole–ring system is delayed. Nevertheless, even in these cases, the ring becomes fragmented relatively quickly due to the combined effects of HD instabilities and efficient radiative cooling that trigger local thermal instabilities. Before this stage, the hole–ring structure preserves a coherent and approximately regular morphology for a limited period, whose duration is highly sensitive to both the FMK properties and the local shell conditions (see Sect.~\ref{sec:sizes}).

\subsection{A hybrid scenario: Late-stage shell puncturing and the origin of quasi-stationary rings}
\label{sec:hybrid}

The morphology of the GM presents a significant physical tension between its thermal properties and its kinematics. On the one hand, the presence of warm dust ($T \approx 150$~K; \citealt{2024ApJ...976L...4D}) and the striking spatial correlation between mid-infrared rings and the "crinkled" X-ray filaments identified by \cite{2024ApJ...964L..11V} argue for shocked CSM. On the other hand, the measured near-zero radial (i.e., along the LoS) velocities ($-50$ to $0$~km~s$^{-1}$; \citealt{2024ApJ...976L...4D}) and the conspicuous lack of bright optical emission (typical of fully shocked dense clumps such as the Quasi-Stationary Flocculi - QSFs) suggest that the bulk of the GM has not yet been significantly accelerated by the forward shock.

As demonstrated by our HD simulations (see Sect.~\ref{sec:shell_evol}), primary FMKs traveling at velocities of $8000-15000$~km~s$^{-1}$ would have penetrated the circumstellar shell a few decades prior to the arrival of the forward shock. This extended duration between the initial puncture and shock impact allows the resulting cavities to enlarge significantly due to lateral expansion. Unless the shell is much denser than current estimates, these primary FMKs would produce holes significantly larger than the $1^{\prime\prime}-3^{\prime\prime}$ structures observed by JWST.

To reconcile the observed scales and the kinematics of the rings, a third possibility would be involving a population of slower "bullets" ($6000-8000$~km~s$^{-1}$) that punctured the shell only a few years before shock impact. These bullets may originate from the secondary fragmentation of RT fingers at the contact discontinuity. \casa\ models that account for radiative losses and non-equilibrium ionization (NEI) effects \citep{2025A&A...696A.188O} demonstrate that RT fingers are highly susceptible to thermal instabilities, particularly given the metal-rich composition of the ejecta. These instabilities fragment the heads of the fingers into discrete, dense clumps that can outpace the average expansion of the main ejecta body, acting as "shrapnels" positioned immediately ahead of the forward shock (e.g., \citealt{2012ApJ...749..156O}). 

Because these secondary bullets travel at more modest velocities and are located in close proximity to the blast wave, they facilitate "late-stage" puncturing of the shell. This minimizes the time available for the holes to expand before the shell is overtaken by the forward shock, thereby maintaining the small, regular ring sizes observed in the GM. This interpretation finds strong observational support in the identification of a "slow" ejecta knot population by \cite{2026arXiv260209175J}, which appears to be a physical extension of the RT finger system (see also \citealt{2023ApJ...953..131K}). 

This hybrid mechanism effectively reconciles the pockmarked morphology with the quasi-stationary kinematics of the rings in the GM. If the interaction between the forward shock and the dense shell is very recent (occurring within the last few years), the bulk of the material would not yet have acquired significant radial momentum. Crucially, in this "first contact" scenario, the forward shock has not yet had sufficient time to disrupt the ring-hole systems; consequently, the structures maintain the circularity and regular symmetry observed by JWST, rather than appearing as the highly distorted filaments typical of long-term shock processing (see Sect.~\ref{sec:shock-shell}). 

At first glance, this framework introduces an apparent fine-tuning problem, as it implies that we are observing the GM at a remarkably specific and fleeting epoch. However, this observational coincidence can be naturally resolved if the surrounding CSM is significantly more structured and inhomogeneous than idealized in single-shell simulations. In a realistic circumstellar environment characterized by multiple shells, clumps, or sheets distributed at varying radial distances from the progenitor, a powerful selection effect emerges. 


Because our simulations demonstrate that well-defined, regular hole-ring systems are intrinsically short-lived (visible only immediately prior to or for a few decades following the forward-shock encounter before being disrupted) the GM may simply represent the only region within the complex CSM architecture where the spatial timing and physical conditions are currently optimal for detection. In other directions, where the shock-shell interaction has either not yet occurred or has already concluded decades ago, the structural remnants of these punctures would be entirely unobservable. Rather than requiring accidental temporal fine-tuning for the entire remnant, the pockmarked appearance of the GM is a spatial selection effect highlighting the exact boundary where the blast wave is currently making first contact with, and slowly accelerating, the progenitor's heterogeneous mass-loss history.

\section{Properties of the holes and rings in the shocked shell and comparison with JWST constraints}
\label{sec:holes_rings}

Across the full set of FMK-driven models, the qualitative morphology of the hole–ring systems is highly consistent. Variations in knot size, velocity, and shell density primarily affect the cavity diameter and rim brightness, whereas the defining configuration (low-density holes surrounded by compressed, ring-like shells) emerges naturally from the FMK–shell interaction. Within this framework, the resulting morphology encodes information on both the dynamical properties of the knots and the local shell density at the time of impact. In principle, a comparison between simulated hole and ring sizes, densities, and evolutionary timescales and those inferred from JWST observations can be used to constrain the properties of the FMKs responsible for the pockmarked structure, as well as those of the surrounding shell.

\begin{figure}
   \centering
   \includegraphics[width=0.48\textwidth]{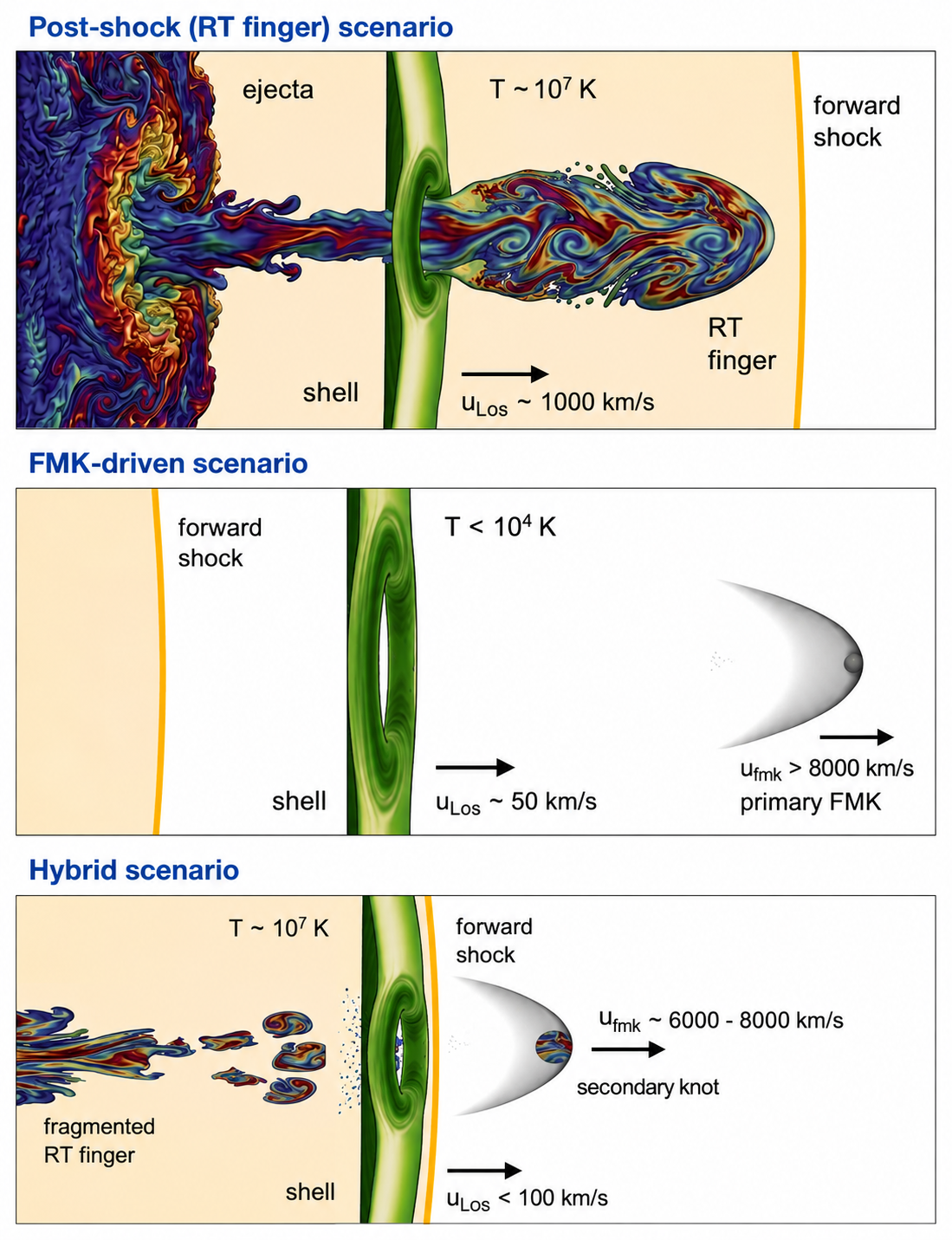}
   \caption{Schematic of the three investigated interaction scenarios detailing the geometry, thermal states, and kinematics of the ring structures. Top panel: Post-shock RT finger scenario (ejecta-driven sculpting), where the forward shock has completely overrun the CSM shell, which is subsequently compressed and penetrated by large-scale RT fingers over long timescales (investigated in \citealt{2025A&A...696A.188O}). Middle panel: Primary FMK-driven scenario (pre-shock puncturing), where high-velocity ejecta knots run ahead of the main blast wave, puncturing the unshocked CSM shell and driving rapid lateral cavity expansion via bow shocks. Bottom panel: Hybrid scenario, where radiatively cooled RT fingers undergo thermal instabilities and fragment behind the blast wave, producing smaller, slower secondary bullets that pierce the shell immediately prior to or during first contact with the forward shock.}   
   \label{schematic_scenarios}%
   \end{figure}

A necessary first step, however, is to establish which of the three proposed formation channels for the GM structure (post-shock sculpting, FMK-driven, or hybrid scenario; see Fig.~\ref{schematic_scenarios}) is the dominant contributor to its observed appearance in \casa. In this section, we therefore compare the hole–ring properties predicted by the FMK-driven scenario with both JWST observations of the GM region and the expectations of the alternative post-shock and hybrid scenarios, with the goal of identifying the mechanism that most plausibly accounts for the observed morphology and of defining observational diagnostics capable of discriminating between the three.

\subsection{Chemical properties of gas filling the holes}

A primary observational diagnostic for discriminating among the proposed scenarios involves analyzing the chemical composition and spatial distribution of the material within the holes. In the post-shock scenario, cavities are created by large-scale, metal-rich ejecta fingers that displace already shocked shell material (see top panel in Fig.~\ref{schematic_scenarios}); consequently, the plasma filling the hole interiors should be dominated by ejecta with enhanced abundances of heavy elements such as O, Si, S, and Fe \citep{2025A&A...696A.188O}. In contrast, in the pure FMK-driven scenario, the holes are excavated long before shock arrival and are subsequently refilled by shocked CSM flowing inward from the cavity boundaries (see middle panel in Fig.~\ref{schematic_scenarios}; see also Figs.~\ref{fmk_evolution} and \ref{hole_disrupt}). In this case, the interior gas should retain a chemical composition nearly identical to that of the circumstellar shell, exhibiting minimal metal enrichment. 

The hybrid scenario involving fragmented RT-fingers introduces an intermediate and highly distinctive signature. While the cavities are excavated within the unshocked CSM (structurally mimicking the FMK framework) the puncturing bullets are themselves metal-rich ejecta fragments (see bottom panel in Fig.~\ref{schematic_scenarios}). Because these fragments pierce the shell only shortly before the forward shock overruns the structure, they are expected to reside either within or immediately adjacent to (i.e., projected toward the observer) the cavities at the current epoch. Under this mechanism, observations should reveal small, localized, high-metallicity ejecta "kernels" embedded within a larger volume of otherwise pristine, shocked CSM.

While JWST imaging is ideally suited to characterize the complex morphology of these hole-ring systems, the plasma filling the cavities is expected to be hot and tenuous, emitting predominantly in the X-ray band. A reliable determination of metal abundances therefore requires a coordinated, multi-wavelength approach: JWST imaging to define the exact cavity boundaries, paired with high-spatial-resolution Chandra X-ray spectroscopy ($\approx 0.5^{\prime\prime}$ on-axis) to probe the chemical composition and ionization age of the interior plasma. 

In practice, however, implementing this diagnostic may be heavily complicated by LoS projection effects, which can artificially superimpose unrelated foreground or background ejecta filaments over a pristine, CSM-filled cavity, as well as by complex HD mixing that can blur these chemical boundaries. Furthermore, a major observational challenge may arise from the intense synchrotron emission produced by the nearby reverse shock, which can contaminate or entirely overwhelm the faint thermal X-ray spectrum of the plasma inside the holes. Disentangling these multi-component contributions and identifying the subtle spectral signatures predicted by the hybrid scenario (namely, a high-temperature CSM component interspersed with small-scale, metal-rich ejecta features) will therefore require highly refined spatial and spectral modeling.

\subsection{Characteristic sizes and evolution timescales of hole–ring systems}
\label{sec:sizes}

\begin{figure*}
   \centering
   \includegraphics[width=0.8\textwidth]{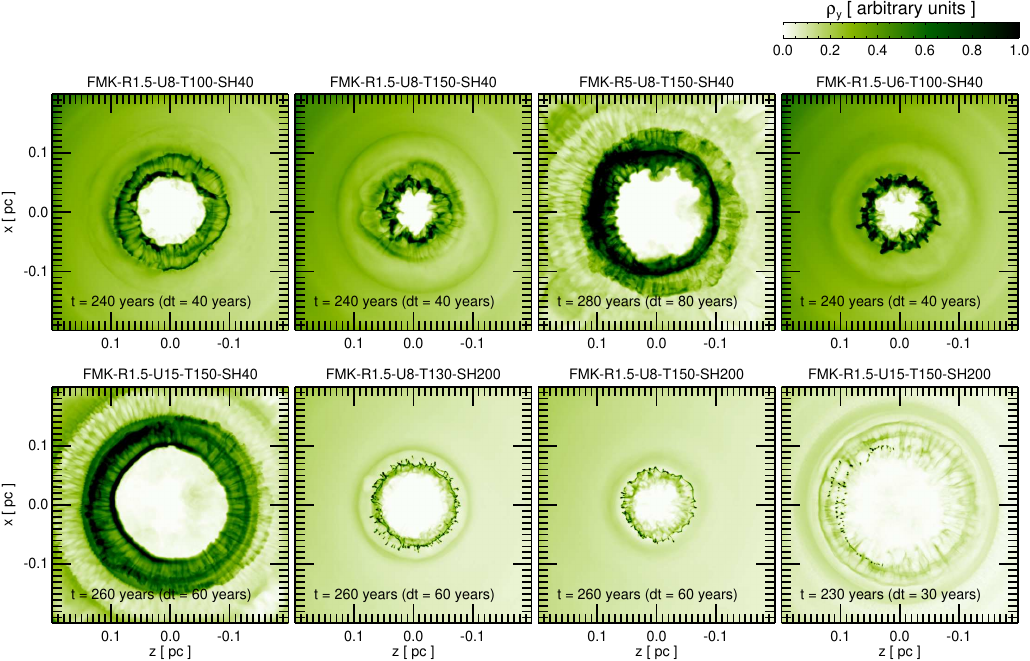}
   \caption{3D volumetric renderings of the shell density in the selected region, viewed from Earth's perspective. Each panel shows one of the models listed in Table~\ref{tab1} at the indicated epoch. \sal{In all models, the forward shock reaches the shell at $t=200$ yr; the value in parentheses gives the time elapsed since the shock impact ($dt$).}}
   \label{holes_rings}%
   \end{figure*}

Another key diagnostic for distinguishing between the formation scenarios is the characteristic size and dynamical evolution of the hole–ring systems. Figure~\ref{holes_rings} illustrates the morphology of the cavity–ring structures across various models between 40 and 80 years after the forward shock encounters the shell. In Appendix~\ref{app:holes_evol}, we present the evolution of these structures for all the models in Table~\ref{tab1} over a 70-year interval immediately following the forward-shock impact.
 
\sal{Before discussing the dependence of the hole-ring properties on the shell density and FMK parameters, we briefly compare their evolution in the proposed formation scenarios. In the post-shock scenario, the rings initially expand as RT fingers penetrate the dense shell, before being disrupted by interactions with neighboring fingers and ejecta clumps \citep{2025A&A...696A.188O}. In the primary FMK-driven scenario, the hole-ring systems typically reach their maximum size before the forward shock arrives, then shrink and are eventually disrupted by HD instabilities, producing relatively large rings. By contrast, in the hybrid scenario, the much shorter interval between shell puncture and shock arrival allows the rings to be observed at different evolutionary stages while they remain relatively small. Thus, the evolution and size of the rings provide additional diagnostics for distinguishing among the proposed formation mechanisms.}

We find that, in general, in models adopting a shell with maximum density $n_{\rm sh}=40$~cm$^{-3}$, the resulting hole–ring systems are substantially larger, even at early times, than the features observed in the GM with JWST, which exhibit typical angular sizes of $1\arcsec-3\arcsec$ \citep{2024ApJ...976L...4D}, corresponding to linear scales of $\sim (5-15)\times 10^{16}$~cm ($0.016-0.048$~pc). This discrepancy suggests that, if the GM has a density comparable to the reference model, the FMKs responsible for the observed structures must be smaller and/or less energetic than those considered here, potentially placing them below current detection limits. Conversely, in models with higher shell densities, such as FMK-R1.5-U8-T120-SH200, FMKs with physical properties consistent with observational constraints produce hole-ring systems that more closely align with the dimensions observed by JWST. However, these features can still become significantly larger than the observed scale, depending on the time interval between the initial puncture and the subsequent shock impact.

Across the full set of models, the smallest cavities produced by FMKs reach characteristic sizes comparable to the upper end of the observed hole distribution, driven by the rapid lateral expansion of the shocked shell following knot passage. We further find that smaller holes and rings at the time of shock impact correspond to shorter observability windows of coherent post-shock hole–ring structures. In low-density shells ($n_{\rm sh}=40$~cm$^{-3}$), well-defined cavities surrounded by thin, bright rims persist for no more than $\sim 30$~yr, while in denser shells ($n_{\rm sh}=200$~cm$^{-3}$) their lifetime extends to at most $\sim 60$~yr (see Fig.~\ref{holes_rings} and Appendix~\ref{app:holes_evol}). Although higher shell densities modestly prolong the survival of FMK-driven structures, they remain intrinsically transient. Conversely, cavities that retain a coherent morphology for longer timescales ($\gtrsim 60$~yr) are produced only in models generating substantially larger holes (e.g., FMK–R1.5–U10–T130, FMK–R1.5–U15–T150, and FMK–R5–U8–T150), with dimensions several times greater than those observed in the GM.

The hybrid scenario, involving fragmented RT fingers, offers a compelling resolution to the size-timescale paradox. In this framework, secondary fragments originating from the contact discontinuity act as "bullets" with relatively modest velocities ($6000-8000$~km~s$^{-1}$). This lower bullet velocity directly results in a reduced lateral expansion rate for the resulting holes. Moreover, because these fragments are located immediately ahead of the blast wave, the time interval between the initial shell puncture and the forward-shock arrival is significantly minimized. This "late-stage puncturing" prevents the cavities from undergoing the prolonged expansion observed in primary FMK models, naturally preserving the small, $1\arcsec-3\arcsec$ diameters characteristic of the GM. Moreover, the rings are not yet distorted by the impact of the forward shock. 

In the hybrid scenario, our simulations predict that the GM will evolve rapidly, with its appearance strongly depending on the time interval between the knot passage and the arrival of the forward shock. If the knot punctures the shell shortly before the shock arrives, the shock disrupts the nascent ring, erasing its coherent circular morphology within a few years. As the forward shock continues to process the dense CSM shell, momentum transfer accelerates the rings, increasing their radial velocities. \sal{Although the hole-ring system ultimately disappears, it leaves behind localized overdensities of shocked shell material (lower-right panel of Fig.~\ref{fmk_evolution}) and a disturbed shell morphology. These overdense relics may persist for decades, or even longer, after the shock-shell interaction. In a GM with a complex multi-layer structure, successive shock interactions with different shells could therefore produce the coexistence of hole-ring systems at different evolutionary stages, together with small-scale overdense relics of earlier interactions. However, as these relics become progressively mixed into the shocked shell, they are unlikely to remain as unambiguous signatures of former FMK interactions.} High-resolution JWST monitoring over the coming years will therefore provide a unique opportunity to witness the transformation of a pristine circumstellar structure into a fully shocked HD complex.

\subsection{Kinematic properties}

The density and kinematic properties of the GM rings offer a critical diagnostic for distinguishing among the three scenarios proposed for their origin. While morphology provides a visual baseline, the velocity field of the material (both in the plane of the sky and along the LoS) reveals the underlying physics of the interaction between high-velocity ejecta and the CSM. In the following we examine whether the tangential velocity magnitude (i.e. in the plane of the sky) carries a distinct signature of the knot-shell interaction, and whether the radial velocity measurements (i.e., along the LoS) from JWST and Chandra can serve as a pivotal benchmark for validating HD models and determining the material's spatial positioning.

   \begin{figure*}
   \centering
   \includegraphics[width=0.8\textwidth]{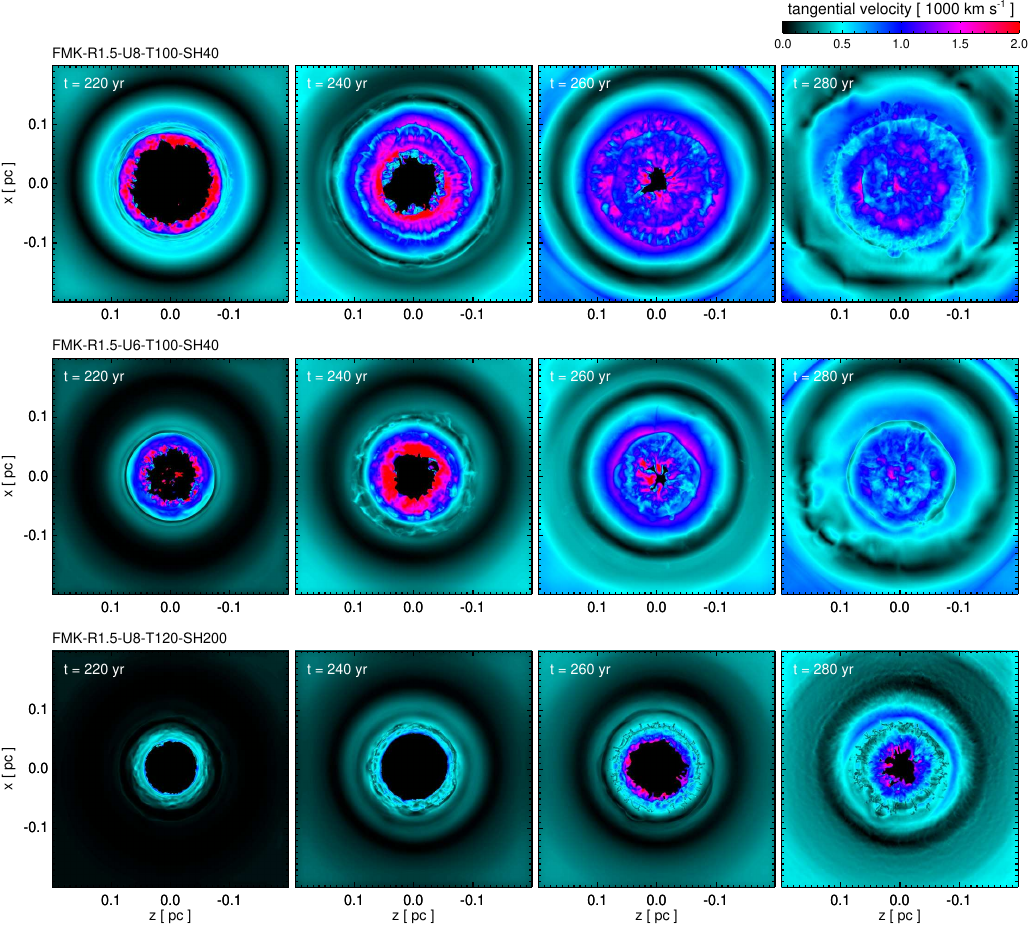}
   \caption{Tangential velocities of the shell material relative to the LoS in models FMK-R1.5-U8-T100-SH40 (top), FMK-R1.5-U6-T100-SH40 (middle), and FMK-R1.5-U8-T120-SH200 (bottom) \sal{at the labeled times.} Velocities are LoS-averaged and weighted by the shell density, considering only cells with densities exceeding 1\% of the shell maximum in the analyzed volume. From left to right, the panels show the evolution of the tangential velocity after the forward shock passage. \sal{The time since shock impact is obtained by subtracting 200 yr from the labeled time.}}
   \label{tang_vel}%
   \end{figure*}

\subsubsection{Tangential velocity}

Figure~\ref{tang_vel} illustrates the tangential velocity magnitude for two models representative of different shell density regimes: FMK-R1.5-U8-T100-SH40 (low shell density) and FMK-R1.5-U8-T120-SH200 (high shell density). Additionally, we include model FMK-R1.5-U6-T100-SH40, which serves as a proxy for the hybrid scenario involving slower, late-arriving bullets. Velocities were averaged along the LoS and weighted by the local shell density, mask-filtering regions where the density falls below 1\% of the pre-shock maximum \citep[following][]{2025A&A...696A.188O}. The black regions in the maps delineate the FMK-carved cavities.

In the low-density shell model ($n_{\rm sh}=40$~cm$^{-3}$), moderate tangential velocities of $\sim 500$~km~s$^{-1}$ are present around the cavities, while the edges of the holes reach $1000-2000$~km~s$^{-1}$, reflecting the violent lateral displacement of material by high-velocity FMKs. Conversely, in the high-density shell model ($n_{\rm sh}=200$~cm$^{-3}$), the transmitted shock propagates more slowly through the medium, producing lower tangential velocities of approximately $500$~km~s$^{-1}$. Peaks exceeding $1500$~km~s$^{-1}$ in the high-density case appear only at later stages ($\sim 60$~yr after shock impact), by which time the hole-ring system is already substantially perturbed by the surrounding environment.

In the hybrid scenario, the kinematic signature is primarily governed by the precise timing of the encounter between the ejecta and the circumstellar shell. If the knot-shell interaction occurs a few years prior to the forward shock's arrival, the lateral expansion velocity and resulting morphology follow the kinematics of the primary FMK-driven scenario; in this case, the puncture grows within a relatively undisturbed, stationary medium before being overtaken by the blast wave, which subsequently disrupts the ring's regularity. Conversely, if the interaction occurs shortly after the shell has been processed by the forward shock, the lateral expansion mimics the dynamics of the post-shock scenario, preserving the circularity of the rings. 

For comparison, post-shock, ejecta-driven sculpting generates tangential velocities of $200-800~\mathrm{km~s^{-1}}$, resulting from the gradual lateral displacement of shocked shell material by dense ejecta "fingers" penetrating the contact discontinuity \citep{2025A&A...696A.188O}. In contrast, primary FMKs impacting low-density shells generate much higher tangential velocities ($1000-2000~\mathrm{km~s^{-1}}$), whereas denser shells suppress the lateral expansion, reducing the velocities to $400-1000~\mathrm{km~s^{-1}}$. The hybrid RT fragmentation scenario is expected to yield similar velocities ($500-1000~\mathrm{km~s^{-1}}$), reflecting the lower speeds of the secondary bullets. Together with the observed small size ($1\arcsec-3\arcsec$), these moderate velocities would support a recent interaction between a dense circumstellar shell and a secondary population of slower ejecta fragments.


\subsubsection{Radial velocity}
\label{sec:radial_velocity}

\begin{figure}
   \centering
   \includegraphics[width=0.45\textwidth]{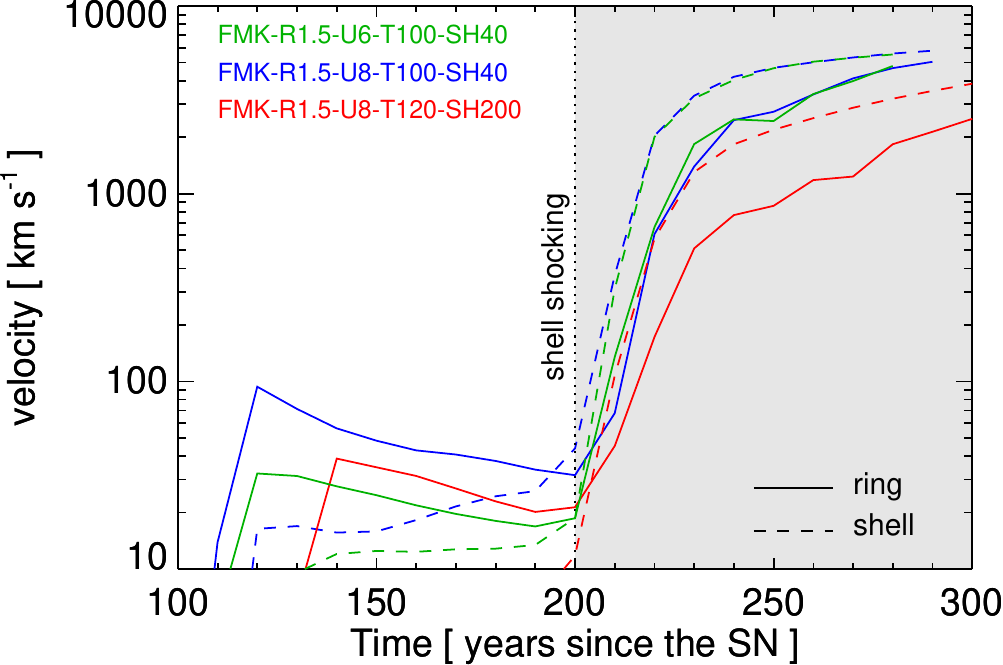}
   \caption{Absolute values of the average radial velocities of the ring material (solid lines) and of the shell material unaffected by the FMK interaction (dashed lines) for models FMK-R1.5-U8-T100-SH40 (green), FMK-R1.5-U6-T100-SH40 (blue), and FMK-R1.5-U8-T120-SH200 (red). The velocities are intrinsically negative, as the material propagates along the negative $y$-axis, i.e., toward the observer. The velocities are averaged along the LoS and weighted by the local shell density. For the ring material, only cells with density exceeding 1\% of that of the unperturbed shell are included. The forward shock of the remnant encounters the pierced shell at an age of approximately 200 years; the gray shaded region in the figure indicates the epoch during which the pierced shell is already shocked.}
   \label{prof_rad_vel}%
   \end{figure}

While tangential velocities trace the expansion of the rings in the plane of the sky, the  LoS velocity measures the rings' radial motion and serves as a direct diagnostic of their acceleration by the SN shock. Because observations place the GM on the nearside of the remnant, positioned in front of the main ejecta shell, any shock-induced acceleration is observed as a blueshift. Figure~\ref{prof_rad_vel} shows the radial velocity of the ring and shell material for models FMK-R1.5-U6-T100-SH40, FMK-R1.5-U8-T100-SH40 and FMK–R1.5–U8–T120–SH200. As for the tangential velocities, the radial velocities were averaged along the LoS and weighted by the local shell density; the ring material has been identified by considering only regions where the density exceeds 1\% of the pre-shock maximum. 

As illustrated in the figure, following the initial impact of the ejecta knot on the circumstellar shell (occurring at approximately 100 and 120 years post-SN for the high-density and low-density shell cases, respectively) a ring-like structure forms and is accelerated to radial velocities in the range $-100$ to $-40$~km~s$^{-1}$ (see solid lines in the figure). The largest blueshifts are found in run FMK-R1.5-U8-T100-SH40, reflecting the reduced resistance to the transverse momentum of the bullet, whereas the lower blueshifts occur in run FMK-R1.5-U6-T100-SH40, consistent with the lower velocity of the bullet penetrating the shell.

\sal{Prior to the arrival of the primary blast wave, the radial velocity of the ring gradually stabilizes between $-40$ and $-20$~km~s$^{-1}$. Upon interaction with the forward shock, the ring is rapidly accelerated toward the observer. The evolution is essentially identical in the two low-density shell models (FMK-R1.5-U6-T100-SH40 and FMK-R1.5-U8-T100-SH40). Figure~\ref{prof_rad_vel} shows that the shocked ring reaches velocities of $\approx -100$~km~s$^{-1}$ within the first decade after shock impact and approaches $\approx -1000$~km~s$^{-1}$ after $20-30$~yr. By contrast, the unperturbed shell material (dashed lines), owing to its lower density, is accelerated more efficiently by the forward shock, reaching $\approx -1000$~km~s$^{-1}$ within $10-20$~yr. In the high-density shell scenario, the acceleration is slower: the shocked ring reaches $\approx -100$~km~s$^{-1}$ after $\sim 20$~yr and $\approx -1000$~km~s$^{-1}$ after $\sim 50$~yr, while the unperturbed shell reaches the same velocities after $\sim 10$ and $\sim 30$~yr, respectively.}


These velocity values can be compared with the LoS velocities inferred from the analysis of JWST and Chandra observations of the GM in \casa. Spectroscopic data from JWST reveal that the bright infrared rings are moving at radial velocities of only $-50$ to 0~km~s$^{-1}$ (\citealt{2024ApJ...976L...4D}). According to our models, this "quasi-stationary" motion indicates that the densest, dust-emitting clumps have experienced minimal radial acceleration despite being processed by the shock (see Fig.~\ref{prof_rad_vel}). This is consistent with the hybrid scenario in which the shell has been pierced recently (keeping the size of the ring-hole system consistent with the dimensions observed) and shocked by no more than one decade to maintain the radial velocity of the rings at the observed levels.

In contrast, X-ray emission associated with the GM exhibits significantly higher acceleration, with blueshifts reaching approximately $-2300$ to $-3500$~km~s$^{-1}$ (\citealt{2024ApJ...964L..11V}). This high velocity is representative of the forward shock speeds measured elsewhere in the remnant, typically ranging between $5000$ to $7000$~km~s$^{-1}$. According to our model, the shell material not affected by the FMK piercing is accelerated more efficiently by the forward shock (because of the lower density), reaching radial velocities between $-1500$ and $-3000$~km~s$^{-1}$ (depending on the model) two to three decades after the interaction with the shock. This implies that the X-ray emission traces lower-density material of the shell unaffected by the piercing compared to the infrared, which is more easily accelerated by the blast wave.

\section{Discussion: The Green Monster as a multiphase, heterogeneous structure}
\label{sec:discussion_multiphase}

The comparison between our HD models and multi-wavelength observations has been complicated by an apparent physical tension between the thermal properties of the GM and its kinematics. On one hand, the presence of warm dust \citep{2024ApJ...976L...4D} and high-temperature plasma ($5 \times 10^6$--$10^7$~K; \citealt{2024ApJ...964L..11V}) suggest a shocked environment. On the other hand, the low radial velocities ($-50$ to $0~\mathrm{km~s^{-1}}$) reported by \citet{2024ApJ...976L...4D} have been used to argue against comprehensive shock processing. 

By integrating recent proper motion data and multi-filter diagnostics, this thermal/kinematic paradox can be naturally resolved without invoking the fine-tuned \sal{timing required by the primary FMK-driven scenario, in which the observed rings would have to be caught during a very short-lived evolutionary phase.} Instead, the observational data point to a highly structured, multiphase CSM. Given the structural complexity and density variations of the circumstellar environment analyzed below, it is highly probable that all three mechanisms (detailed schematically in Fig.~\ref{schematic_scenarios}) coexist and operate simultaneously within the GM. Acting in tandem, these scenarios carve out morphological features at different spatial scales, locations, and evolutionary epochs.

   \begin{figure*}
   \centering
   \includegraphics[width=0.8\textwidth]{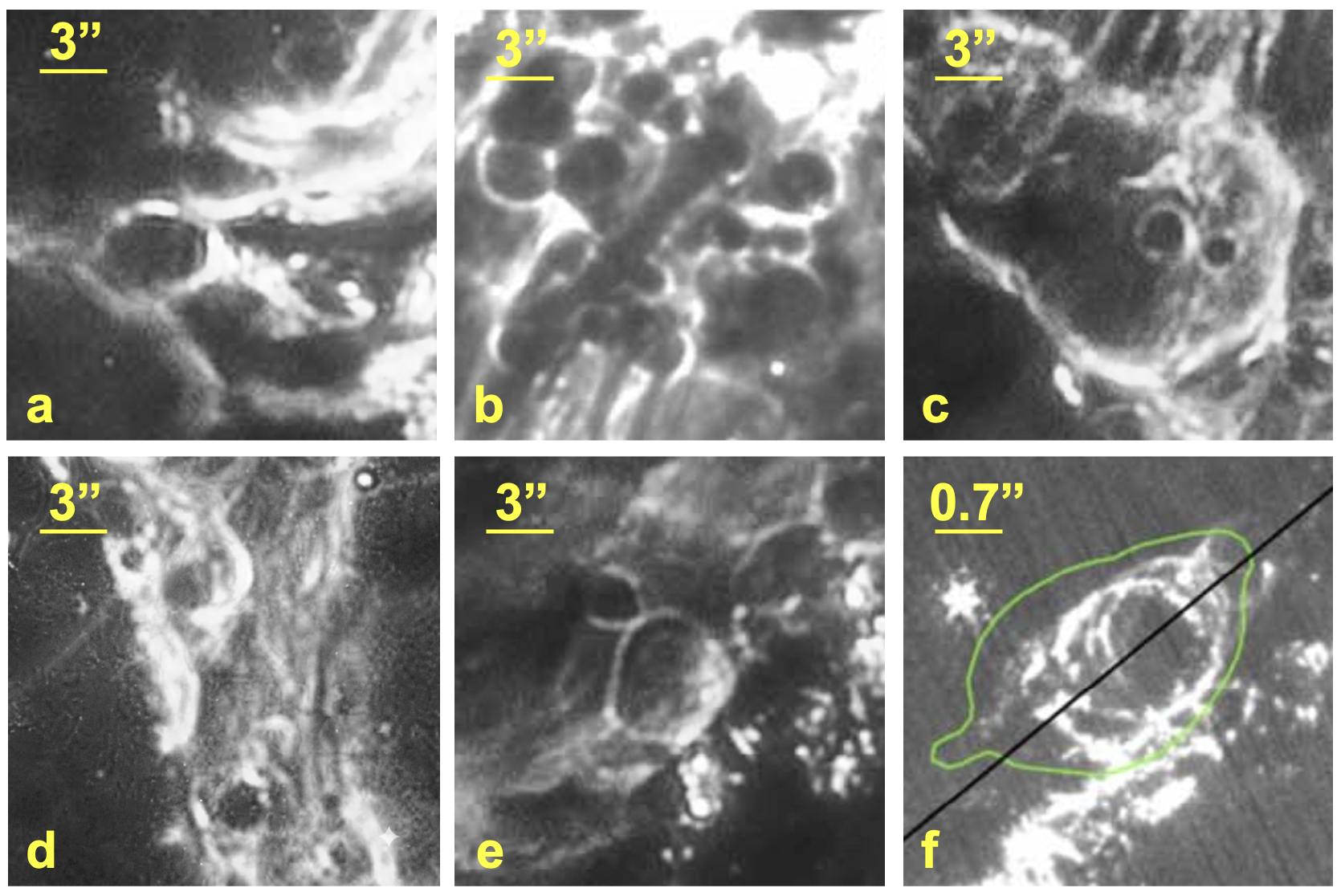}
   \caption{Examples of hole-ring systems observed with JWST (panels adapted from \citealt{2024ApJ...976L...4D}), illustrating features caught at different evolutionary stages and embedded within environments of highly non-uniform density. These snapshots suggest that hole-ring structures produced by distinct physical mechanisms may simultaneously coexist within the GM: features driven by the post-shock sculpting scenario (e.g., panel b), large-scale cavities originating from the primary FMK-driven scenario (such as the extended loop visible in panel c or the large ring in panel e), and compact features matching the predictions of the hybrid scenario (the small inner rings nested within the large loop in panel c). Panel f shows an example of a multi-ring structure detected at position (d3) in Fig.~4 of \citet{2024ApJ...976L...4D}, providing compelling evidence for a multi-layered or stratified shell structure where cavities form at different epochs depending on the arrival time of ejecta knots at each specific layer.}
   \label{jwst_gallery}%
   \end{figure*}

\subsection{Selection effects and the multi-density shell}

The exceptionally low radial velocities isolated by \citet{2024ApJ...976L...4D} were obtained primarily from specific hole-ring structures prominent in the JWST F162M filter. This narrow-band filter is heavily dominated by $[\mathrm{Fe~II}]$ line emission, which selectively traces slow, dense, radiative shocks ($100-500~\mathrm{km~s^{-1}}$). Crucially, the majority of the hole-ring networks visible in mid-infrared (MIR) continuum maps are either faint or entirely absent in the F162M band. 

This behavior implies a profound selection effect. Rather than being a homogeneous layer, the GM is most likely composed of dense regions and filaments embedded within a significantly more diffuse, ambient shell spanning a broad range of densities. A compilation of selected hole-ring systems observed with JWST highlights this structural diversity (see Fig.~\ref{jwst_gallery}), displaying features that appear to be caught in completely different evolutionary stages and embedded within environments of highly non-uniform density. While the diffuse component may have a baseline density close to the nominal value adopted in our reference model ($n_{\mathrm{sh}} \approx 40~\mathrm{cm^{-3}}$), the local structures illuminated in $[\mathrm{Fe~II}]$ may correspond to very dense regions. These features likely possess pre-shock densities on the order of $n_{\mathrm{sh}} \approx 100-1000~\mathrm{cm^{-3}}$, mimicking the properties of QSFs (\citealt{2023ApJ...953..131K}). Furthermore, this complex configuration is likely exacerbated by a multi-layered CSM architecture, where multiple concentric or overlapping shells reside at different radial distances from the progenitor star (see panel f in Fig.~\ref{jwst_gallery}).

Our HD simulations show that when the SNR forward shock encounters a ring structure of high density, the acceleration timescale is noticeably prolonged. The material experiences an immediate thermal shock, lighting up in radiative diagnostics, but its heavy inertia keeps its radial velocity near zero ($\lesssim 100~\mathrm{km~s^{-1}}$) for a few decades post-impact (see Fig.~\ref{prof_rad_vel}). Conversely, in the diffuse regions ($n_{\mathrm{sh}} \approx 40~\mathrm{cm^{-3}}$), the shock propagates much faster as a non-radiative blast wave ($v_{\mathrm{shock}} > 500~\mathrm{km~s^{-1}}$). 

This multi-density and multi-shell architecture may be rigorously tested through the analysis of real-time structural variations caught by a multi-year baseline comparison of JWST imaging. Under this hypothesis, multiple rings embedded within the lower-density rings of the GM ($n_{\mathrm{sh}} \approx 40~\mathrm{cm^{-3}}$) or belonging to more distant, recently reached shell layers should exhibit detectable long-term signatures of rapid fading, changing brightness gradients, or complete structural disintegration. Because our models show that hole-ring systems in low-density environments are highly volatile and easily distorted after being overrun by the blast wave, any broken arcs, incomplete morphologies, or missing sections identified across a 3-year baseline would represent regions where the forward shock has already processed and dismantled the diffuse components of the shell structure. 

Conversely, the pristine, circular rings visible in Fig.~\ref{jwst_gallery} (panels a, b, and d, and the two nested rings in panel c) would correspond exclusively to the heavy, high-inertia clumps ($n_{\mathrm{sh}} \gtrsim 200-1000~\mathrm{cm^{-3}}$) caught during their initial, unperturbed phase of shock interaction. In these dense substructures, or in internal shell layers where the shock interaction has only just begun, the prolonged acceleration timescale preserves their geometric regularity against the advancing blast wave over short observational baselines.

\subsection{Large- and small-scale rings}

A particularly intriguing structural signature in the GM is the discovery of large, extended filamentary loops (such as the prominent circular structure tracing a radius of $\approx 6^{\prime\prime}$, corresponding to $\sim 0.1$~pc; see panel c in Fig.~\ref{jwst_gallery}) which enclose smaller, pristine hole-ring systems. Within our modeling framework, this nested morphology can be interpreted as a temporal sequence of distinct ejecta-shell interactions. The large outer loop is highly consistent with our primary FMK models, where a high-momentum knot has punctured the unshocked shell long ago, leaving a massive cavity that has expanded laterally over many decades, thus being partly distorted and refilling the cavity with less dense CSM material. The smaller rings nested inside would then reflect a second phase of "late-stage" puncturing driven by slower, secondary RT fragments or localized post-shock sculpting.

However, a critical alternative is that these large loops do not trace a single coherent structure, but are instead projection effects along the LoS, created by the superposition of unrelated dense filaments in a stratified, multi-layered CSM sheet. Resolving this topological ambiguity through deep 3D kinematic mapping is essential, as the presence of true nested rings would directly validate a continuous, multi-scale projectile population originating from the SN blast.

\section{Summary and conclusions}
\label{sec:summary}

We have investigated the FMK-driven scenario for the formation of the GM in \casa\ using 3D HD simulations of individual FMKs interacting with a dense circumstellar shell, followed by the impact of the remnant’s forward shock. Our results, combined with previous studies of post-shock sculpting \citep{2025A&A...696A.188O}, provide a framework to assess the origin of the GM’s pockmarked morphology and to identify observational diagnostics capable of distinguishing between these scenarios. Our key conclusions are as follows:

\begin{itemize}
\item Mechanism of formation. In the FMK-driven (pre-shock) scenario, holes are carved by bow shocks ahead of supersonically moving knots penetrating the unshocked shell. In the post-shock scenario, cavities are created by the displacement of already shocked shell material by RT fingers and clumps. We introduced here a third, hybrid scenario involving the secondary fragmentation of RT fingers into dense, metal-rich clumps \citep{2012ApJ...749..156O, 2025A&A...696A.188O}. These fragments act as slower bullets ($6000-8000$~km~s$^{-1}$) that puncture the shell immediately before or during the forward shock impact. All these mechanisms may operate simultaneously in \casa.

\item Morphology and lifetime of hole–ring systems. FMK-driven holes and rings exhibit a consistent qualitative morphology: low-density cavities surrounded by compressed rings emerge naturally from the FMK–shell interaction, and the size and brightness of the rims depend on knot properties and local shell density. However, these structures are inherently transient (in the post-shock scenario, the resulting rings and holes continue to evolve, remaining relatively long-lived but progressively growing and merging; see \citealt{2025A&A...696A.188O}). In our FMK simulations, regular hole–ring systems persist for only $\sim 30-60$ years after the forward shock passes. In the hybrid scenario these structures are smaller in comparison with the FMK-driven scenario because the interaction with the shell occurred a few years before the shell gets shocked shortly before the present observations. Holes in denser shells survive somewhat longer, but still collapse and fragment on timescales of the order of $\approx 60$~years. 

\item Scale mismatch and the hybrid solution. Primary FMKs (velocity $> 8000$~km~s$^{-1}$) typically generate cavities larger than the $1\arcsec$–$3\arcsec$ structures observed by JWST \citep{2024ApJ...976L...4D}. This discrepancy arises from their high velocities, which drive rapidly expanding lateral shocks in the shell, combined with the long delay between the initial puncture and the arrival of the main blast wave. The hybrid scenario naturally resolves this tension. In this case, secondary RT fragments are slower and impact the shell at later times, leaving insufficient time for the resulting cavities to expand significantly in the lateral direction before the shock impact. This mechanism reproduces the observed spatial scales and implies a very recent "first contact" between the shock and the GM.

\item Gas Composition. In the FMK-driven scenario, holes are refilled by shocked circumstellar material, and thus the gas within the cavities is expected to have a composition close to the local CSM, with comparatively low metal enrichment. In contrast, the post-shock scenario predicts that cavities are dominated by metal-rich ejecta (enhanced abundances of O, Si, S, Fe) displaced from the contact discontinuity. The hybrid scenario predicts an intermediate signature: a cavity primarily filled with recently shocked CSM, potentially containing localized, high-metallicity "kernels" from the RT-fragments. Combined JWST imaging (to locate and measure the hole–ring systems) and X-ray spectroscopy (e.g., with Chandra, to probe the chemical composition of the hot plasma inside the cavities) may provide a direct, multi-wavelength diagnostic to distinguish between the two formation mechanisms.

\item Kinematic signatures. The tangential and radial velocity fields of the GM rings provide a robust framework for distinguishing between the three scenarios investigated. In the plane of the sky, tangential velocities around FMK-driven holes in low-density shells can reach $1000-2000$~km~s$^{-1}$. In the hybrid and post-shock scenarios, these velocities are more modest ($200-1000$~km~s$^{-1}$), reflecting a less violent displacement of shell material. This lower velocity range is more consistent with the observed lack of bright optical emission. Along the LoS, the radial velocity serves as a direct diagnostic of the rings' acceleration by the remnant's forward shock. Our models indicate that after an initial FMK impact, a ring-like structure is accelerated to radial velocities between $-100$ and $-40$~km~s$^{-1}$, stabilizing between $-40$ and $-20$~km~s$^{-1}$ prior to the arrival of the primary blast wave. The current "quasi-stationary" radial velocities of $-50$ to 0~km~s$^{-1}$ observed by JWST suggest that the dust-emitting clumps have experienced minimal acceleration, consistent with a hybrid scenario in which the shell was pierced recently and has been shocked for no more than two decades. In contrast, the unperturbed shell material is accelerated much more efficiently by the forward shock, reaching the high blueshifted velocities consistent with the $-2300$ to $-3500$~km~s$^{-1}$ detected in X-ray observations.
\end{itemize}

The transient behavior of primary FMK-driven holes contrasts, in principle, with the well-defined, circular rings of relatively small size observed in the GM. In our simulations, sharply bounded cavities and coherent rings with dimensions comparable to those seen by JWST are most prominent only shortly before, or within a few decades ($\sim 30-60$~yr) after, the forward shock passes through the shell. If the shell were entirely shocked at an age of $\sim 200$~yr, as modeled by \citet{2022A&A...666A...2O} to reproduce the global properties of the reverse shock, it would be challenging to reconcile the persistence of regular hole-ring systems with the $\sim 150$~yr elapsed since the forward-shock impact. 

On the other hand, the post-shock scenario faces significant difficulties in explaining the kinematic properties of the GM. In a framework where the rings are formed by the interaction of ejecta fingers with a long-shocked shell, the momentum transfer by the forward shock and the underlying flow of the shocked plasma would inevitably accelerate the rings to radial velocities of thousands of $\mathrm{km~s^{-1}}$. This stands in stark contradiction to the quasi-stationary radial velocities ($-50$ to $0~\mathrm{km~s^{-1}}$) measured by \citet{2024ApJ...976L...4D}.

The hybrid scenario involving fragmented RT fingers provides a robust solution to these chronological and kinematic discrepancies. In this framework, the puncturing "bullets" are secondary clumps that outpace the main ejecta body and strike the shell immediately before or during its interaction with the forward shock. This "first contact" occurs much more recently than the large-scale shell interaction modeled in \citet{2022A&A...666A...2O}. Because the blast wave has only begun processing these specific sub-structures within the last few years, the hole-ring systems have not yet had time to acquire significant radial momentum, nor have they been subject to the long-term HD instabilities that would otherwise disrupt their circularity. 

At first glance, this framework introduces an apparent fine-tuning problem, as it implies we are observing the GM at a remarkably specific and fleeting epoch. However, this observational coincidence is naturally resolved within a highly structured, multiphase CSM. Crucially, as shown in Sect.~\ref{sec:radial_velocity}, the prolonged acceleration timescale of these ultra-dense rings means they can remain quasi-stationary for decades even after being overrun by the forward shock. It is not merely that the shock front hit them recently; rather, their massive inertia makes them highly resistant to picking up radial momentum from the blast wave. Well-defined hole-ring networks thus become a spatial selection effect, highlighting the exact boundaries where the blast wave is currently making first contact with the progenitor's heterogeneous mass-loss history.

The physical challenges encountered here in reconciling the GM's kinematics are directly analogous to those frequently faced in strongly interacting extragalactic SNe (such as Type IIn). In those distant, unresolved systems, mass-loss rates inferred from X-ray, radio, and optical observations commonly produce discrepant values (e.g., \citealt{2025ApJ...984...71D, 2025ApJ...985...51N, 2024Natur.627..759Z}), reflecting how different multi-wavelength diagnostics selectively probe distinct components of a clumpy, overlapping CSM (a representative configuration of which has been inferred for SN 2014c; e.g., \citealt{2024ApJ...977..118O}). 

In this light, our simulations establish the GM as a promising, spatially resolved local laboratory for the effects of a heterogeneous CSM. They demonstrate how a highly inhomogeneous, multi-density CSM naturally yields coexisting, widely disparate kinematic and thermal features via selection effects. While diffuse regions of the shell are processed rapidly, the ultra-dense rings act as high-inertia anchors that preserve their pristine, circular geometry during first contact.

A multi-wavelength observational approach, combining JWST infrared imaging to map the hole-ring structures with Chandra X-ray spectroscopy to probe the composition and kinematics of the plasma inside the cavities, offers the most promising strategy to definitively map these coexisting scenarios across the GM. Future monitoring of the GM's evolution with JWST will be crucial; our model predicts a rapid transition from circular regularity to fragmented filaments and a measurable increase in radial (LoS) expansion velocities as the forward shock begins to fully dismantle the clumpy components of the shell over the coming decades.

\begin{acknowledgements}
We are grateful to Gabriel Rigon for his insightful comments and suggestions, which helped improve the manuscript. The \PLUTO\ code is developed at the Turin Astronomical Observatory (Italy) in collaboration with the Department of General Physics of  Turin University (Italy) and the SCAI Department of CINECA (Italy). We acknowledge the CINECA ISCRA Award N.HP10BUMIQR for the availability of HPC resources and support at the infrastructure Galileo100 based in Italy at CINECA. Additional computations were carried out on the HPC system MEUSA at the SCAN (Sistema di Calcolo per l'Astrofisica Numerica) facility for HPC at INAF-Osservatorio Astronomico di Palermo. 
S.O., M.M., and F.B. acknowledge financial contribution from the PRIN 2022 (20224MNC5A) - ``Life, death and after-death of massive stars'' funded by European Union – Next Generation EU, and the INAF Theory Grant ``Supernova remnants as probes for the structure and mass-loss history of the progenitor systems''.
At Garching, support by the German Research Foundation (DFG) through the Collaborative Research Centre ``Neutrinos and Dark Matter in Astro- and Particle Physics (NDM),'' Grant SFB-1258-283604770, and under Germany's Excellence Strategy through the Cluster of Excellence ORIGINS EXC-2094-390783311 is acknowledged.
F.B. acknowledge support of INAF grant GO-GTO2024.
\end{acknowledgements}


\bibliographystyle{aa}
\bibliography{references}

\clearpage
\begin{appendix}
\onecolumn

\section{Model implementation with the PLUTO code}
\label{app:code}

All simulations were performed with the \textsc{pluto} code, a modular Godunov-type framework for astrophysical fluid dynamics \citep{2012ApJS..198....7M}. We adopt the linearized Roe Riemann solver, which provides an accurate description of shock dynamics through characteristic decomposition of the Roe matrix. Time integration is carried out using a third-order Runge-Kutta (\textsc{RK3}) scheme in an unsplit formulation. Combined with the Roe solver, this yields a numerically robust and minimally diffusive scheme. A monotonized-central flux limiter (\textsc{MC\_LIM}) is applied to the primitive variables to suppress spurious oscillations near strong shocks while maintaining high spatial accuracy.

The code has been extended with additional modules to treat: (i) non-equilibration of electron and ion temperatures, and (ii) deviations from ionization equilibrium. Electron heating at collisionless shocks is implemented assuming rapid energization to $kT_{\mathrm e}\approx0.3$~keV via lower-hybrid waves \citep{2007ApJ...654L..69G}, followed by Coulomb equilibration in the post-shock plasma \citep[see][]{2015ApJ...810..168O}. Nonequilibrium ionization states for the most abundant species are tracked using the ionization age parameter, computed self-consistently during the remnant evolution \citep{2015ApJ...810..168O}.

\sal{It is worth noting that the shock transition itself is not kinetically resolved. The simulations treat shocks as HD discontinuities, while the post-shock electron temperature is initialized through a parameterized prescription accounting for collisionless electron heating (\citealt{2015ApJ...810..168O, 2016ApJ...822...22O, 2020A&A...636A..22O}), followed by Coulomb equilibration and non-equilibrium ionization. This approximation primarily affects the thermal plasma properties and synthetic emission, whereas the large-scale dynamics of the FMK-shell interaction are controlled by the fluid momentum and density evolution and are therefore expected to be largely insensitive to the microphysics operating within the unresolved shock transition.}

To trace the evolution of the shell material, we introduce a passive scalar tracer, $C_{\rm sh}$, associated with the shell. To this end, we solve an additional continuity equation for the tracer, coupled self-consistently to the standard set of HD equations. The quantity $C_{\rm sh}$ represents the mass fraction of shell material within each computational cell. It is initialized as $C_{\rm sh} = 1$ in the shell and $C_{\rm sh} = 0$ in the ambient medium. This approach allows us to follow the mixing of materials during both the FMK–shell interaction and the subsequent propagation of the forward shock, which naturally produces regions with $0 < C_{\rm sh} < 1$. At any time $t$, the local density of shell material in a given cell is then given by $\rho_{\rm sh} = \rho\, C_{\rm sh}$, where $\rho$ is the total mass density. This formulation enables us to isolate the contribution of the shell and to reconstruct the time-dependent morphology of the shocked shell.
\sal{Note that the passive scalar is employed solely as a diagnostic tracer to distinguish material originating from the circumstellar shell and to visualize its morphology. It is not used to determine the shell position or influence the HD evolution. Although numerical diffusion in the Eulerian scheme slightly broadens the scalar across contact discontinuities, this occurs over only a few computational cells and has no impact on the shell dynamics or on our analysis, provided the simulation has sufficient resolution in general.}

The initial setup corresponds to the dashed box schematically shown in Fig.~\ref{fig1}. The computational domain samples a portion of the circumstellar shell and includes an isolated FMK located ahead of the forward shock. In model W15-IIb-sh-MHD+dec-rl-hr of \citet{2025A&A...696A.188O}, the $y$-axis is defined along the LoS, with the observer (Earth) located at negative $y$ values. To remain consistent with this reference frame, and thereby enable a direct comparison with our previous post-shock study \citep{2025A&A...696A.188O} as well as with the observed geometry of the GM (which lies on the near side of the remnant close to its projected center), we adopted a configuration in which the FMK propagates along the $y$-axis toward decreasing $y$. This choice ensures that the simulated kinematics can be interpreted under the same viewing orientation used in the earlier work. The 3D domain is discretized on a uniform Cartesian grid with $512\times1280\times512$ cells, extending from $-1.2$ to $-2.2$~pc along the $y$-axis and from $-0.2$ to $0.2$~pc along both the $x$- and $z$-axes. This corresponds to a uniform spatial resolution of $\Delta x = \Delta y = \Delta z \approx 8\times10^{-4}$~pc ($\approx 2.4\times10^{15}$~cm).

\section{Hole-ring system evolution}
\label{app:holes_evol}

   \begin{figure*}
   \centering
   \includegraphics[width=0.92\textwidth]{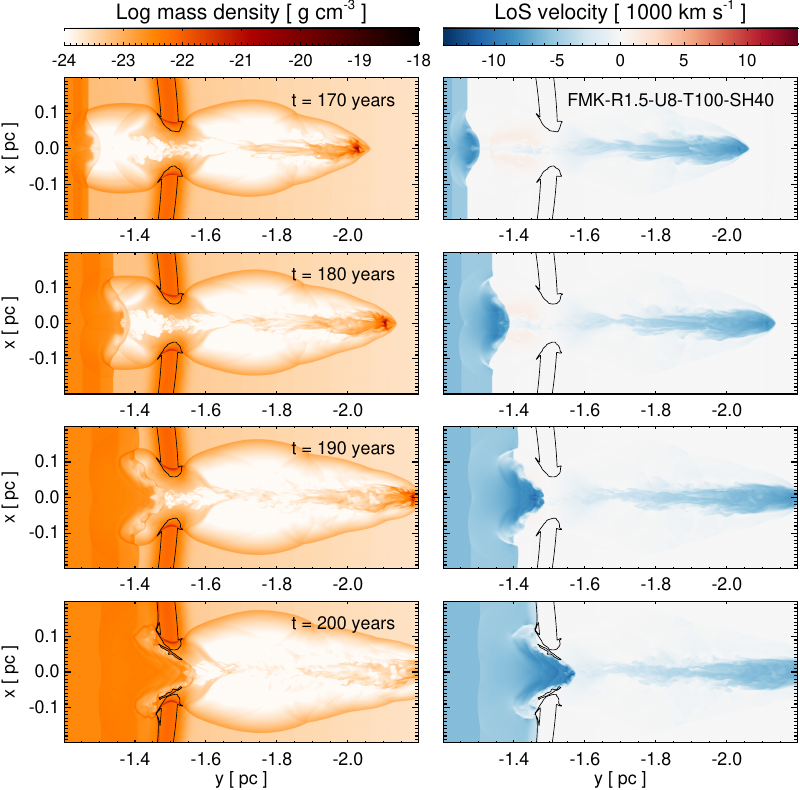}
   \caption{Similar to Fig.~\ref{fmk_strcuct}, but showing the propagation of the remnant's forward shock immediately before its interaction with the circumstellar shell in model FMK-R1.5-U8-T100-SH40. The forward shock is locally distorted as it propagates through the low-density wake left by the FMK. The times are indicated in the upper-right corner of the left panels; the time to shock impact is obtained by subtracting 200 yr from the labeled time.}
   \label{fig:pre_impact}%
   \end{figure*}

   \begin{figure*}
   \centering
   \includegraphics[width=0.92\textwidth]{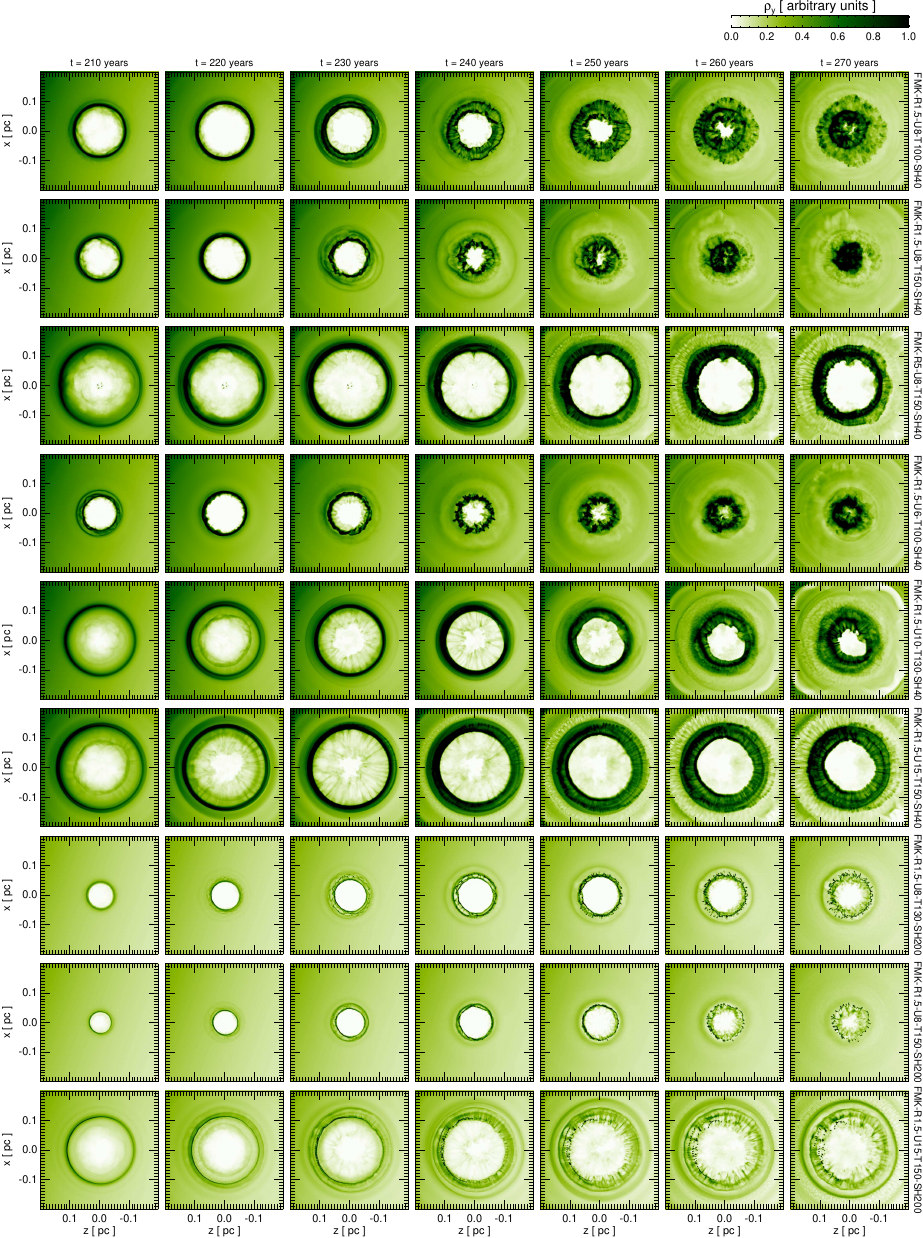}
   \caption{Evolution of hole–ring structures in all FMK-driven simulations (see Table~\ref{tab1}). Each panel shows a 3D volumetric rendering of the shell density in the region of interest, viewed from Earth’s perspective. Each row corresponds to one of the models listed in Table~\ref{tab1}, while panels from left to right show the temporal evolution in 10-year intervals immediately following the impact of the forward shock on the circumstellar shell (see labels on the top of the figure)\sal{; the time since shock impact is obtained by subtracting 200 yr from the labeled time.} This figure illustrates the transient nature of FMK-driven hole–ring systems and allows direct comparison with the sizes and morphologies of holes observed by JWST.}
   \label{holes_evol}%
   \end{figure*}

In this appendix, we provide a detailed description of the temporal evolution of the hole–ring systems in all FMK-driven simulations presented in this work. These models explore a range of knot sizes, velocities, and shell densities (see Table~\ref{tab1}) to encompass the diversity of conditions relevant for \casa. Our aim is to illustrate how the morphological, kinematic, and density properties of the holes and surrounding rings evolve in response to different FMK and shell parameters.

\sal{Figure \ref{fig:pre_impact} illustrates the morphology of the forward shock immediately before it encounters the circumstellar shell. A localized deformation is visible where the shock propagates through the low-density wake excavated by the FMK. The reduced ambient density causes a local increase in the shock velocity, producing a slight protrusion of the shock front. The deformation is generated self-consistently by the HD evolution and reflects the spatial extent of the FMK wake rather than any imposed perturbation of the ambient medium. We note, however, that the extent of the bow-shock wake is limited by the size of the computational domain ($[-1.2, -2.2]$~pc along the $y$-axis). In reality, before the FMK reaches the shell, its bow shock is expected to evolve over a substantially longer distance (especially in the FMK-driven scenario), allowing the wake to become more extended and the remnant shock to develop a more pronounced corrugation by the time it encounters the shell than in our simulations. Such a more structured shock front would likely interact with the shell in a more complex manner, promoting stronger shear flows and HD instabilities during the impact. Consequently, the shocked rings may lose their regular morphology more rapidly than predicted here, suggesting that our estimates of the survival time of coherent hole–ring structures are conservative.}

Across all models, the qualitative morphology of the hole–ring system is highly consistent: low-density cavities are carved by the bow shock preceding the FMK, surrounded by compressed, ring-like rims. Variations in knot size, velocity, or shell density primarily affect the cavity diameter, ring thickness, and rim brightness (see Fig.~\ref{holes_evol}). The lifetime of well-defined structures scales with the initial size of the cavity at the time of forward-shock impact and with the local shell density. In general:

\begin{itemize}

\item Smaller or slower knots produce more compact cavities with thin, bright rims, but these structures are transient, typically persisting for only $\sim 30-60$~yr after the forward shock passes (e.g., models FMK-R1.5-U8-T150-SH40, FMK-R1.5-U6-T100-SH40, and FMK-R1.5-U8-T150-SH200).

\item Larger or faster knots generate larger holes, thicker rings, and stronger compression in the shell, extending the observable lifetime of the cavity up to more than $\sim 80$~yr (e.g., models FMK-R5-U8-T150-SH40, FMK-R1.5-U10-T130-SH40, FMK-R1.5-U15-T150-SH40, and FMK-R1.5-U8-T120-SH200). However, these structures quickly grow beyond the sizes observed by JWST unless the knots are unusually small or slow. In addition the rings appear highly perturbed and the holes filled with low density filaments (see panels in rows 3, 5, 6, and 9 in Fig.~\ref{holes_evol}).

\item Models with intermediate knot sizes and velocities (e.g., models FMK-R1.5-U8-T100-SH40, FMK-R1.5-U8-T120-SH200) show proportional behavior: cavity size and ring thickness increase with knot momentum, while lifetime scales roughly with initial cavity size and local shell density. Thinner or less dense shells reduce both the lifetime and coherence of the rings.

\item Shell density strongly influences both the size and persistence of the holes: denser shells slow the propagation of the transmitted shock, reducing lateral expansion and prolonging the survival of coherent rings, while low-density shells allow rapid cavity expansion and early disruption of the rims.

\end{itemize}

These results highlight the transient nature of FMK-driven structures. While FMKs can generate holes and rims qualitatively similar to the GM, their survival times are limited. This provides a clear framework to compare simulated structures with JWST observations and to constrain the relative timing of FMK and shock interaction with the shell as well as the properties of both the knots and the local circumstellar shell.

\end{appendix}

\end{document}